\documentclass[preprint,tightenlines,aps,showpacs,letterpaper]{revtex4}

\usepackage{graphicx}
\usepackage{slashed}
\usepackage{comment}

\usepackage{amssymb}

\usepackage{amsmath}
\usepackage{amssymb}
\usepackage[hidelinks]{hyperref}
\usepackage{tikz}
\usepackage{tikz-feynman}

\usepackage{color}

\usepackage{xspace}

\newcommand{\notE}{ \hbox{{$E$}\kern-.60em\hbox{/}}}
\newcommand{\notp}{\ \hbox{{$p$}\kern-.43em\hbox{/}}}

\newcommand{\tauvis}{\ensuremath{\tau_{\text{vis}}}\xspace}
\newcommand{\taureco}{\ensuremath{\tau_{\text{reco}}}\xspace}

\preprint{\font\fortssbx=cmssbx10 scaled \magstep2
\hbox to \hsize{
\hskip1.2in 
\hbox{\fortssbx The University of Oklahoma}
\hskip0.2in $\vcenter{
  \hbox{\bf arXiv: [hep-ph]}
  \hbox{\bf OU-HEP-251108}
  \hbox{November 2025}}$ }
}

\begin{document}

\title{\vspace*{0.7in}
Charming top interactions with a neutral Higgs Boson decaying into
$\tau^+ \tau^-$ at the LHC}

\author{
Chenyu Fang\footnote{E-mail address: chenyu.fang@ou.edu},
Phillip Gutierrez\footnote{E-mail address: pgutierrez@ou.edu} and
Chung Kao\footnote{E-mail address: Chung.Kao@ou.edu}
}

\affiliation{
Homer L. Dodge Department of Physics and Astronomy,
University of Oklahoma, Norman, OK 73019, USA }

\date{\today}

\bigskip

\begin{abstract}
  
We investigate the discovery potential of flavor changing neutral
Higgs (FCNH) interactions in top quark decays at the LHC.
A general two Higgs doublet model (2HDM) is adopted to study the top quark
decay $t \to c \phi^0$,
where $\phi^0$ is either a CP-even heavy neutral scalar ($H^0$), 
or a CP-odd pseudoscalar ($A^0$), followed by the Higgs boson
decaying into $\tau^+\tau^-$.
The complete process is
$pp \to t\bar{t} \to t c\phi^0 \to (bjj) (c\tau_{lep}\tau_{had}) +X$,
where $t$ or $\bar{t}$ become virtual when $m_{\phi} > m_t -m_c$, 
$\tau_{lep}$ and $\tau_{had}$ represent the
$\tau$ leptonic decay ($\tau^\pm \to \ell^\pm +\slashed{E}_T$) and the
$\tau$ hadronic decay ($\tau^\pm \to j_\tau +\slashed{E}_T$), respectively.   
We evaluate the cross section applying realistic selection criteria
for the FCNH signal and the dominant physics background at the parton-level.
In addition, we employ \textsc{MadGraph}, \textsc{Pythia}~8, and
\textsc{Delphes} to perform a realistic event-level Monte Carlo simulation
analysis.
The collinear approximation is applied since the $\tau$'s from 
Higgs boson decays are highly boosted such that $\tau_{lep}$ and $\tau_{had}$ 
momenta are approximately in the same direction as the $\tau$'s,
which enables the reconstruction of the Higgs boson mass.
The energy of the charm quark 
provides an additional
kinematic variable that discriminates the signal from the background.
The measured properties of the 125 GeV Higgs boson 
at the LHC implies the alignment limit for 2HDMs
($\lambda_{tch} \propto \cos(\beta-\alpha) \approx 0$).
This means the FCNH couplings for the heavy Higgs bosons ($H^0$ and $A^0$)
will not be suppressed [$\lambda_{tcH} \propto \sin(\beta-\alpha) \approx 1$].
We study the discovery potential for the heavy neutral Higgs bosons at the
LHC with collider energies of $\sqrt{s} = 13$ TeV, 13.6 TeV and 14 TeV,
and find 
the High-Luminosity LHC offers great promise for the discovery of 
heavier Higgs bosons from $t \to c \phi^0$ or $t^* \to c \phi^0$. 

\end{abstract}

\maketitle

\section{Introduction}

The discovery of a 125 GeV CP-even scalar Higgs boson ($h^0$)
in 2012~\cite{Aad:2012tfa,Chatrchyan:2012ufa}
completed the particle content of the Standard Model (SM).
However, it cannot explain some deviations from the SM, such as
the baryon asymmetry in the Universe~\cite{Sakharov:1967dj}.
Possible solutions can be derived from
two Higgs doublet models (2MDMs), which are simple and natural extensions to the
Standard Model. They provide possible explanations for why the top quark is so much
heavier than other elementary fermions~\cite{Das:1995df}, 
why the up quark is so
much lighter than charm and top quark~\cite{Xu:1991gs}, and provide additional sources of CP
violation~\cite{Liu:1987ng, Weinberg:1990me, Fuyuto:2017ewj}
to explain the observed baryon asymmetry of the Universe.

General 2HDMs lead to a rich phenomenology at the CERN Large Hadron Collider
(LHC), including a Higgs
pseudoscalar ($A^0$), a pair of charged Higgs bosons ($H^\pm$), and
a heavier Higgs scalar ($H^0$). Furthermore, there could be
significant flavor changing neutral currents mediated by the CP-even
heavy scalar 
and the CP-odd pseudoscalar, 
as well as
enhanced charged Higgs decay ($H^+ \to c \bar{b}$) without CKM
($V_{cb}$) suppression~\cite{Ghosh:2019exx}. 
The SM expectation is ${\cal B}(t \to c h^0) \approx 10^{-14}$~\cite{
 AguilarSaavedra:2004wm,Mele:1998ag,Eilam:1990zc}, which is significantly
less than current and near term experiments can observe.
If this FCNH signal is observed at the LHC,
it would imply physics beyond the Standard Model~\cite{
Hou:1991un,AguilarSaavedra:2000aj,Kao:2011aa,Chen:2013qta,Atwood:2013ica,
Khachatryan:2014jya,Durieux:2014xla,Chen:2015nta,Khachatryan:2016atv,
Aaboud:2018oqm,Papaefstathiou:2017xuv,Jain:2019ebq,Arroyo-Urena:2019qhl,
Castro:2020sba,Zhang:2020naz,ATLAS:2024mih,CMS:2024ubt}.

We adopt the Yukawa Lagrangian in a general
two Higgs doublet model~\cite{Davidson:2005cw,Mahmoudi:2009zx} as 
\begin{equation}\label{eq:yukawaL}
  \begin{aligned}
{\cal L}_Y =& \frac{-1}{\sqrt{2}} \sum_{\scalebox{0.6}{F=U,D,L}}
 \bar{F}\Big\{  \left[ \kappa^Fs_{\beta-\alpha}+\rho^F c_{\beta-\alpha} \right] h^0 +
   \left[ \kappa^Fc_{\beta-\alpha}-\rho^Fs_{\beta-\alpha} \right] H^0 \\
   &- i \, {\rm sgn}(Q_F)\rho^F A^0 \Big\} P_R F 
  -\bar{U} \left[ V \rho^D P_R - \rho^{U\dagger} V P_L \right] D H^+
  -\bar{\nu} \left[ \rho^L P_R \right] L H^+ + {\rm H.c.} \, 
\end{aligned}
\end{equation}
where $P_{L,R} \equiv ( 1\mp \gamma_5 )/2$,
$c_{\beta-\alpha} \equiv \cos(\beta-\alpha)$,
$s_{\beta-\alpha} \equiv \sin(\beta-\alpha)$,
$\alpha$ is the mixing angle between neutral Higgs scalars,
$\tan\beta \equiv v_2/v_1$ is the ratio of
the vacuum expectation values of the two Higgs doublets~\cite{Gunion:1989we},
$Q_F$ is the fermion charge, and the
$\kappa$~matrices are diagonal and fixed by
fermion masses to $\kappa^F = \sqrt{2}m_F/v$ with $v \approx 246$~GeV, while
the $\rho$ matrices contain both diagonal and off-diagonal
elements with free parameters.
In addition, $F,U,D,L$ represent the elementary fermions,
up-type quarks, down-type
quarks, and charged leptons, respectively. The matrix elements $\rho$
are the FCNH couplings to the fermions.
The consistency between experimental data and the Standard
Model~\cite{Sirunyan:2018koj,Aad:2019mbh},
implies that all two Higgs doublet models must be in 
the alignment limit~\cite{Craig:2013hca,Carena:2013ooa}
with one SM-like 125 GeV light scalar ($h^0$).

According to recent searches for $t \to c h^0$ with
the Higgs boson decaying into final states with leptons as well as 
$h^0 \to b\bar{b}$ and $h^0 \to \gamma\gamma$, 
the ATLAS Collaboration~\cite{ATLAS:2024mih},
and the CMS Collaboration~\cite{CMS:2024ubt} 
placed a strong constraint on
the branching fraction ${\cal B}(t \to c h^0) \leq 3.7 \times
10^{-4}$, 
which leads to an upper limit on the FCNH Yukawa coupling $|\lambda_{tch}|$
\begin{equation}
    \lambda_{tch} \leq 0.036 \, ,
\end{equation}
with the relation between $\lambda_{tch}$ and
the $t \to c h^0$ branching
fraction~\cite{TheATLAScollaboration:2013nbo} being
\begin{equation}
    \lambda_{tch} \approx 1.92\times\sqrt{\mathcal{B}(t \to c h^0)} \, ,
\end{equation}
$|\lambda_{tch}| = \tilde{\rho}_{tc}\cos(\beta-\alpha)$, and
\begin{equation}
  \tilde{\rho}_{tc} \approx \sqrt{(|\rho_{tc}|^2 + |\rho_{ct}|^2)/2}
\end{equation}
where the $\rho$ matrix can be chosen to be non-Hermitian such that
$|\rho_{ct}| < |\rho_{tc}|$.
Therefore, the effective Lagrangian for $H^0$ and $A^0$ can be written as,
\begin{equation}
  \mathcal{L} = \frac{\lambda_{tcH}}{\sqrt{2}}\bar{c}t H^0
              +i\frac{\lambda_{tcA}}{\sqrt{2}}\bar{c}t A^0 + H.c.
  \label{eq:effLag}
\end{equation}
where
\begin{equation}
  |\lambda_{tcH}| = \tilde{\rho}_{tc}\sin(\beta-\alpha)
  \quad {\rm and} \quad 
  |\lambda_{tcA}| = \tilde{\rho}_{tc} \, 
\end{equation}
are not suppressed in the alignment limit.

In a previous study~\cite{Gutierrez:2020eby},
we demonstrated that employing the collinear approximation~\cite{Ellis:1987xu,Rainwater:1998kj,Plehn:1999xi} for
$h^0 \to \tau\tau \to \ell^+\ell^- +\slashed{E}_T,\, \ell^\pm = e^\pm, \mu^\pm$,
in charming top decays of $t \to c h^0$,
we can reconstruct the Higgs and the top quark effectively 
at the LHC.
In this article, we investigate the discovery potential of
$H^0$ or
$A^0$ through 
flavor changing top quark decays ($t \to c \phi^0$)
followed by the Higgs boson decaying into $\tau^+ \tau^-$ with
one $\tau$ decaying leptonically and the other decaying hadronically
($\tau^+\tau^- \to \ell^\pm j_\tau +\slashed{E}_T$)
at the LHC,
where $\slashed{E}_T$ is the missing transverse energy in the event
from the neutrinos. 

We perform a Monte Carlo simulation using
\textsc{MadGraph}~5~\cite{Alwall:2011uj},
\textsc{Pythia}~8~\cite{Sjostrand:2014zea} and
\textsc{Delphes}~\cite{deFavereau:2013fsa}, and compare with a
parton level analysis to study the FCNH decay of one top quark
while the other top quark decays hadronically to a bottom quark ($b$) and 
two light jets: 
$pp \to t\bar{t} \to b W^{\pm} c \phi^0 \to b j j c \tau^+ \tau^- +X$,
where $X$ represents the other particles produced in
the $pp$ collision.
We have calculated the production rates using the full tree level matrix
elements including the Breit-Wigner resonance for both the signal and
the background processes.

Since we did not apply charm tagging, our analysis is suitable
for a general search for $t \to q h^0, q = u, c$. Many previous
studies have adopted the Cheng-Sher Ansatz~\cite{Cheng:1987rs}
as the benchmark Yukawa coupling 
\begin{equation}
  \lambda_{tqh} = \frac{\sqrt{2 m_t m_q}}{v}
\end{equation}  
where $q = u, c$ and $v \approx 246$~GeV is the Higgs vacuum
expectation value. The FCNH coupling, as the geometric mean for top
and charm quark masses, is
\begin{equation}
  \lambda_{tch} = \frac{\sqrt{2 m_t m_c}}{v} \approx 0.09 \, ,
\end{equation}
which has been excluded by recent ATLAS and CMS
measurements~\cite{ATLAS:2024mih,CMS:2024ubt}.
For simplicity, we assume $\lambda_{tch} \gg \lambda_{tuh}$ and focus
on the search for $t \to c h^0$.

There are several aspects to note in this analysis.
 To reconstruct the Higgs boson
 and the real top quark, the collinear approximation of
 tau decays~\cite{Hagiwara:1989fn,Hagiwara:2012vz} is used.
 The collinear approximation for reconstructed tau decays are
 required to have a momentum
 fractions $x_i$ ($0 < x_i < 1$),
 where $x_i = p({\tauvis}_i)/p(\tau_{\text{reco}}), i = 1, 2$,
 with ${\tauvis}_{i}$ representing the visible component of the tau decay,
 either a light-lepton or jet, and $\taureco$ the reconstructed tau lepton,
 more effectively reduced the physics background than the centrality
 requirement suggested in Refs.~\cite{Chen:2015nta,Aaboud:2018oqm}.
 Furthermore, the energy of the charm quark in the top quark rest frame
 provides good acceptance for the FCNH
 top quark signal while rejecting background~\cite{Kao:2011aa,Han:2001ap}.
 Promising results are presented for the LHC with $\sqrt{s} = 14$ TeV.

\section{Higgs Signal and Event Selections}

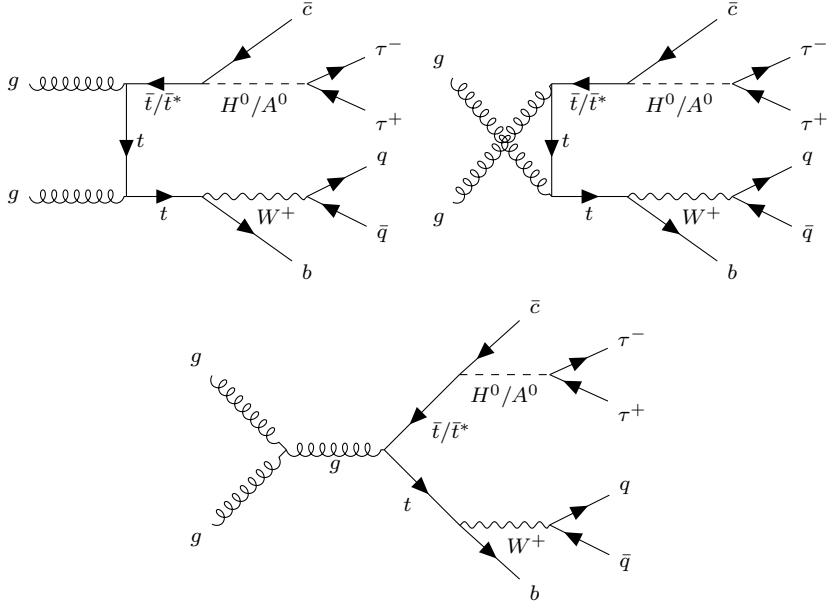
\begin{figure}
\centering
\begin{tikzpicture}[font=\scriptsize]
  \begin{feynman}
    \vertex (a) at (0,-1.5) {\(g\)};
    \vertex (b) at (0,0) {\(g\)};
    \vertex (c) at (1.5,-1.5);
    \vertex (d) at (1.5,0);
    \vertex (e) at (2.5,-1.5);
    \vertex (f) at (2.5,0);
    \vertex (g) at (3.9,-1.5);
    \vertex (h) at (4.9,-1) {\(q\)};
    \vertex (i) at (4.9,-2) {\(\bar{q}\)};
    \vertex (j) at (3.9,-2.5) {\(b\)};
    \vertex (k) at (3.9,0);
    \vertex (l) at (3.9,1) {\(\bar{c}\)};
    \vertex (m) at (5.0,0.5) {\(\tau^-\)};
    \vertex (n) at (5.0,-0.5) {\(\tau^+\)};

    \diagram* {
      (a) -- [gluon] (c), 
      (c) -- [anti fermion, edge label'=\(t\)] (d), 
      (b) -- [gluon] (d),  
      (c) -- [fermion, edge label'=\(t\)] (e),  
      (d) -- [anti fermion, edge label'=\(\bar{t}/\bar{t}^*\)] (f), 
      (e) -- [boson, edge label'=\(\quad\quad W^+\)] (g), 
      (f) -- [scalar, edge label'=\( H^0/A^0\)] (k), 
      (g) -- [fermion] (h), 
      (g) -- [anti fermion] (i), 
      (e) -- [fermion] (j), 
      (f) -- [anti fermion] (l), 
      (k) -- [fermion] (m), 
      (k) -- [anti fermion] (n), 
    };
  \end{feynman}
\end{tikzpicture}
\begin{tikzpicture}[font=\scriptsize]
  \begin{feynman}
    \vertex (a) at (0,0.3) {\(g\)};
    \vertex (b) at (0,-1.8) {\(g\)};
    \vertex (c) at (1.5,-1.5);
    \vertex (d) at (1.5,0);
    \vertex (e) at (2.5,-1.5);
    \vertex (f) at (2.5,0);
    \vertex (g) at (3.9,-1.5);
    \vertex (h) at (4.9,-1) {\(q\)};
    \vertex (i) at (4.9,-2) {\(\bar{q}\)};
    \vertex (j) at (3.9,-2.5) {\(b\)};
    \vertex (k) at (3.9,0);
    \vertex (l) at (3.9,1) {\(\bar{c}\)};
    \vertex (m) at (5.0,0.5) {\(\tau^-\)};
    \vertex (n) at (5.0,-0.5) {\(\tau^+\)};

    \diagram* {
      (a) -- [gluon] (c), 
      (c) -- [anti fermion, edge label'=\(t\)] (d), 
      (b) -- [gluon] (d),  
      (c) -- [fermion, edge label'=\(t\)] (e),  
      (d) -- [anti fermion, edge label'=\(\bar{t}/\bar{t}^*\)] (f), 
      (e) -- [boson, edge label'=\(\quad\quad W^+\)] (g), 
      (f) -- [scalar, edge label'=\( H^0/A^0\)] (k), 
      (g) -- [fermion] (h), 
      (g) -- [anti fermion] (i), 
      (e) -- [fermion] (j), 
      (f) -- [anti fermion] (l), 
      (k) -- [fermion] (m), 
      (k) -- [anti fermion] (n), 
    };
  \end{feynman}
\end{tikzpicture}
\begin{tikzpicture}[font=\scriptsize]
  \begin{feynman}
    \vertex (a) at (0,-1.2) {\(g\)};
    \vertex (b) at (0,1.2) {\(g\)};
    \vertex (c) at (1.2,0);
    \vertex (d) at (2.5,0);
    \vertex (e) at (3.5,-1);
    \vertex (f) at (3.5,1);
    \vertex (g) at (4.7,-1);
    \vertex (h) at (5.7,-0.5) {\(q\)};
    \vertex (i) at (5.7,-1.5) {\(\bar{q}\)};
    \vertex (j) at (4.5,-1.9) {\(b\)};
    \vertex (k) at (4.7,1);
    \vertex (l) at (4.5,1.9) {\(\bar{c}\)};
    \vertex (m) at (5.8,1.5) {\(\tau^-\)};
    \vertex (n) at (5.8,0.5) {\(\tau^+\)};

    \diagram* {
      (a) -- [gluon] (c), 
      (c) -- [gluon, edge label'=\(g\)] (d), 
      (b) -- [gluon] (c),  
      (d) -- [fermion, edge label'=\(t\)] (e),  
      (d) -- [anti fermion, edge label'=\(\bar{t}/\bar{t}^*\)] (f), 
      (e) -- [boson, edge label'=\(\quad\quad W^+\)] (g), 
      (f) -- [scalar, edge label'=\( H^0/A^0\)] (k), 
      (g) -- [fermion] (h), 
      (g) -- [anti fermion] (i), 
      (e) -- [fermion] (j), 
      (f) -- [anti fermion] (l), 
      (k) -- [fermion] (m), 
      (k) -- [anti fermion] (n), 
    };
  \end{feynman}
\end{tikzpicture}
\caption{Leading-order Feynman diagrams for the signal.}
\label{fig:FeynmanDiagram}
\end{figure}

This section presents the cross section and search strategy for the
FCNH signal
$pp \to t \bar{t} \to bjj \bar{c} \phi^0 \to bjj \bar{c} \tau^-\tau^+
+X$ at the LHC.
We focus on the discovery channel with one $\tau$ decaying hadronically
($\tau \to j_\tau \nu_\tau, j_\tau = \pi, \rho$, or $a_1$),
while the other tau lepton decays leptonically into a light-lepton ($\mu^\pm$ or $e^\pm$)
and neutrinos.
Therefore, the final state of our FCNH signal is
$pp \to t\bar{t} \to (bjj) (\bar{c} \ell^\mp j_\tau)+\slashed{E}_T +X$.
The leading-order Feynman diagrams are presented in
Figure~\ref{fig:FeynmanDiagram}.

For our study, the matrix elements for 
$gg, q\bar{q} \to t\bar{t} \to b jj c\phi^0 \to bjj c\tau^+\tau^-$
are generated using \textsc{MadGraph}~5~\cite{Alwall:2011uj} at LO.
In addition, we apply the collinear approximation for
tau decays~\cite{Hagiwara:1989fn,Hagiwara:2012vz},  
$\phi^0 \to \tau^+ \tau^- \to j_\tau \ell +\slashed{E}_T$. 
Since the Higgs boson is much more massive than the tau lepton
($m_{\phi} \gg m_\tau$), the taus from the Higgs decay are highly boosted.
To a good approximation, the $j_\tau, \ell^\pm$, and neutrinos 
travel in the same direction as the tau lepton.
Thus, we can reconstruct the Higgs boson mass from the invariant mass
of the tau lepton pairs ($M_{\tau\tau}$).

The parton level cross section is evaluated using the 
\textsc{CT14LO} parton distribution functions (PDFs)~\cite{Dulat:2015mca}.
For simplicity, the factorization scale ($\mu_F$) and
the renormalization scale ($\mu_R$) are  chosen to be the invariant mass of
the top quark pair ($M_{t\bar{t}}$).
With the above scale choices and PDFs, the $K$-factors are computed using TOP++~\cite{Czakon:2011xx}. 
For all three LHC energies, $\sqrt{s} = 13,\, 13.6,\, \text{and}\, 14$~TeV, the $K$-factors for top quark pair production are found to be approximately the same, around 1.8.

In the event level analysis, the parton level signal samples are generated
using \textsc{MadGraph}~\cite{Alwall:2011uj,Alwall:2014hca,Frederix:2018nkq}
with \textsc{2HDM}~\cite{Degrande:2014vpa} model.
Then the sample is processed by \textsc{Pythia}~8~\cite{Sjostrand:2014zea},
to simulate the parton showering and hadronization, and the $\tau$ 
decay by \textsc{Tauola}~\cite{Jadach:1990mz,Jadach:1993hs}.
Finally, detector simulation is applied to the events using \textsc{Delphes}~\cite{deFavereau:2013fsa}.

To provide a realistic estimate for production rates at the LHC,
we evaluate the cross section for the FCNH signal and physics
background in $pp$ collisions with 
the proper tagging and mistagging efficiencies.
The ATLAS $b$-tagging efficiencies~\cite{Aad:2019aic}
are adopted to evaluate the cross section for the FCNH signal and
physics background. 
The $b$-tagging efficiency is 0.7,
the probability that a $c$-jet is mistagged as a $b$-jet ($\varepsilon_c$)
is approximately 0.14, while
the probability that a light jet ($u, d, s, g$) is mistagged
as a $b$-jet ($\varepsilon_j$) is 0.01.
For tau-jets ($j_\tau$), we apply the ATLAS tagging efficiency $\varepsilon_\tau = 0.75\,(0.60)$ and mistagging efficiencies $\varepsilon_\tau(j) = 0.03\,(0.002)$ for 1-prong (3-prong) tau-jets, respectively, following Refs.~\cite{ATLAS:2015xbi,ATLAS:2019uhp}.

\subsection{Event Selections}

From the discussion of the signal in the previous section,
every event is required to contain at least five jets, including exactly
one identified $b$ jet, one identified $\tau$ jet, and one light-lepton.
We require events satisfy the following requirements, which are
similar to the ATLAS and CMS $h^0 \to
\tau^+\tau^-$ studies~\cite{Aaboud:2018pen}.
\begin{itemize}
\item[(i)] At least five jets with $p_T \geq 25$ GeV and $|\eta| \leq 2.5$.
  including one $b$-jet and one $\tau$-jet, 
\item[(ii)] One lepton with $p_T \geq 20$ GeV and $|\eta| \leq$2.5.
\item[(iii)] An angular separation ($\Delta R=\sqrt{(\Delta\phi)^2+(\Delta\eta)^2}$) greater than 0.4
  is required between every pair of particles.
\item[(iv)] Missing transverse energy $\slashed{E}_T \geq 30$ GeV.
\end{itemize}

Two light jets $j_1$ and $j_2$, are selected by minimizing $|M_{jj}-m_W|$
and $|M_{bjj}-m_t|$.
If the Higgs boson is lighter than the top quark ($m_{\phi} < m_t$),
the $c$-jet is selected by minimizing $|M_{c\tau\tau}-m_t|$, where
the $\tau$ leptons are reconstructed by applying the collinear approximation.

\begin{figure}[htb]
 \centering
 \includegraphics[width=68mm]{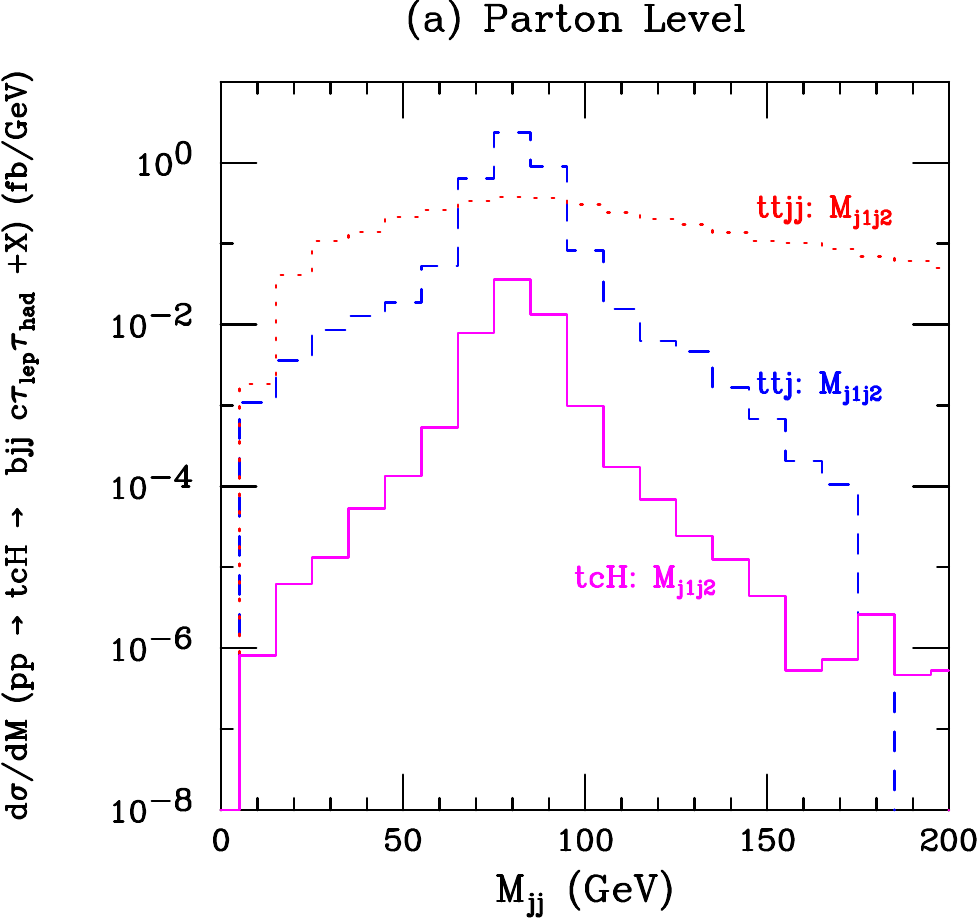}
   \hspace{1mm}
 \includegraphics[width=68mm]{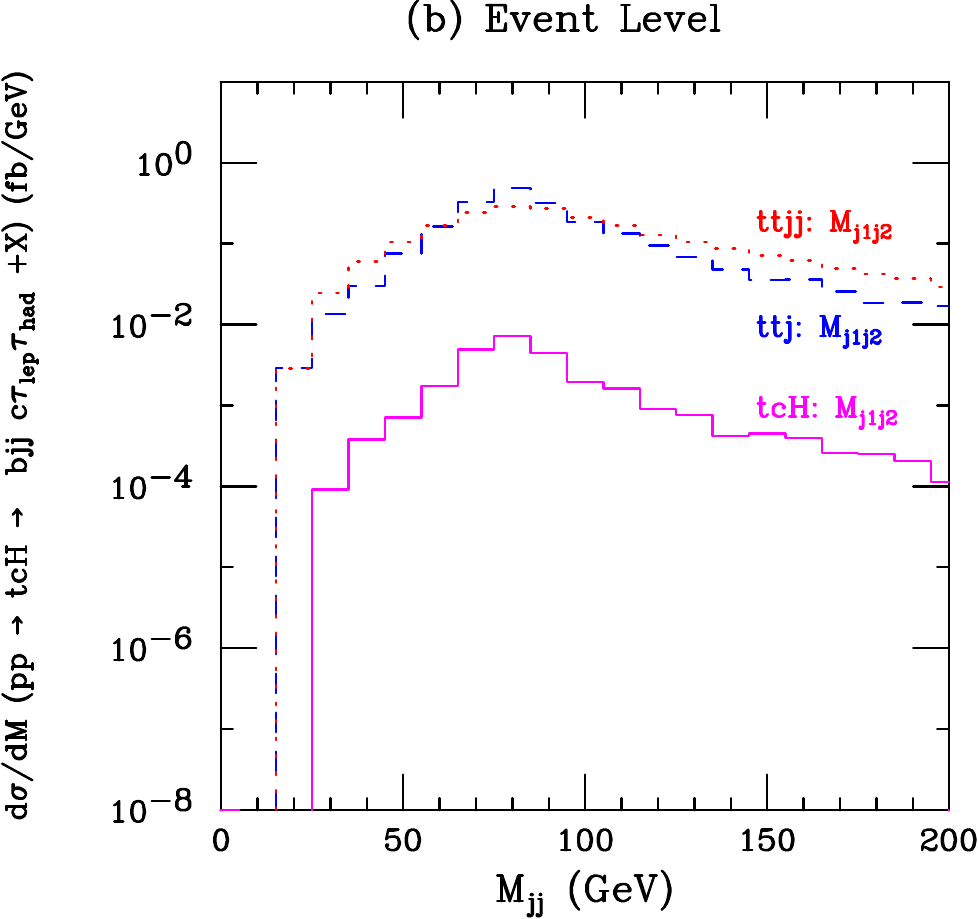}
 \caption{Invariant mass distribution $d\sigma/dM_{jj}$ for the Higgs
   signal (magenta, solid) and for the dominant
     background from $ttj$ (blue, dashed) and $ttjj$ (red, dotted) at the LHC with $\sqrt{s} = 13$
     TeV. Results are shown for (a) parton level,
     and (b) event level.}
   \label{mjj}
\end{figure}

\begin{figure}[htb]
 \centering
 \includegraphics[width=68mm]{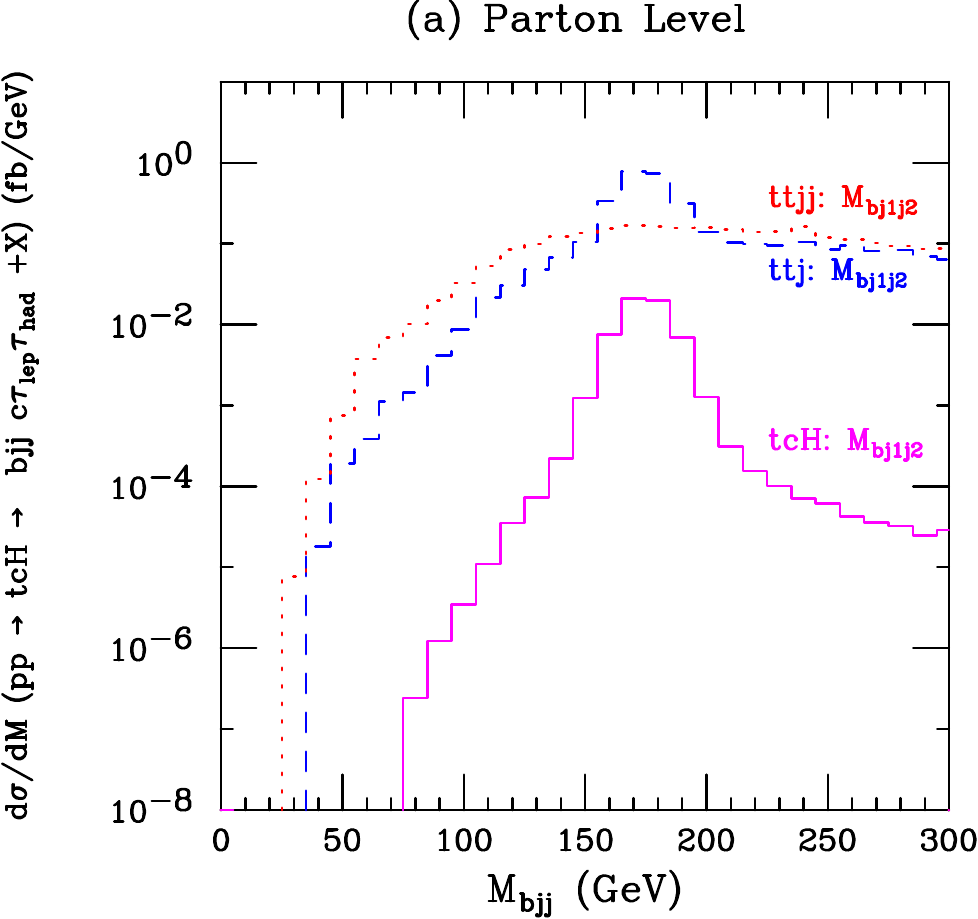}
   \hspace{1mm}
 \includegraphics[width=68mm]{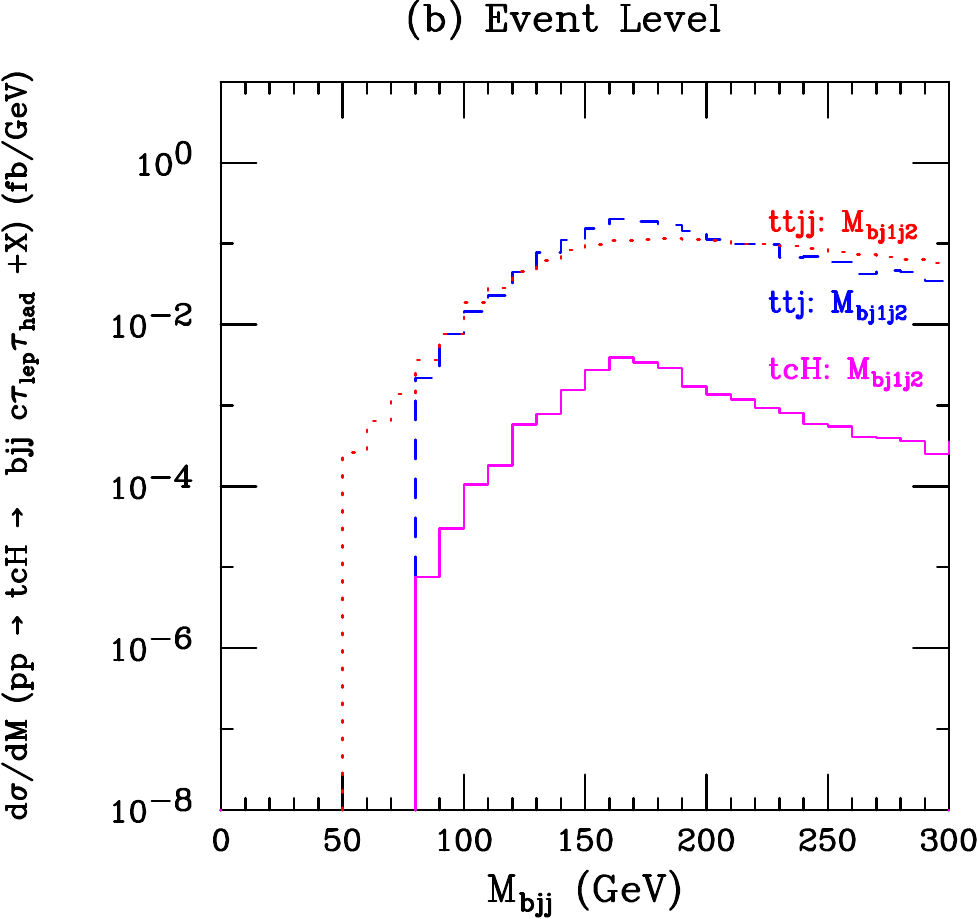}
 \caption{Invariant mass distribution $d\sigma/dM_{bjj}$ for the Higgs
   signal (magenta, solid) and for the dominant
     background from $ttj$ (blue, dashed) and $ttjj$ (red, dotted) at the LHC with $\sqrt{s} = 13$ TeV.
     Results are shown for (a) parton level, and (b) event level.}
   \label{mbj1j2}
\end{figure}

Since $j_1$ and $j_2$ result from $W$ boson decay,
their invariant mass distribution, $M_{j_1 j_2}$, peaks at
$m_W \approx 80.4$~GeV.
Additionally, one top quark decays into a $W$ boson and a $b$ quark, 
thus $M_{bj_1 j_2}$ has a peak at $m_t \approx 172.7$ GeV.
Figures~\ref{mjj} and \ref{mbj1j2} present the invariant mass
distribution of $M_{j_1 j_2}$ and $M_{bj_1 j_2}$, respectively, at both
the parton and event level.
Using the ATLAS mass resolution~\cite{Aad:2019mkw},
we require the reconstructed $W$ and top quark masses lie in the mass windows 
$\Delta M_{j_1 j_2} = 0.20 M_W$ and
$\Delta M_{b j_1 j_2} = 0.25 m_t$. 

\subsection{Higgs Mass Reconstruction}

For the FCNH signal,
$t \to c\, \phi^0 \to c\, \tau^+ \tau^- \to c\, \ell^\pm j_\tau +\slashed{E}_T$,
we consider two situation:
(a) for $m_{\phi} < 150$ GeV, the invariant mass $M_{\tau^+\tau^-}$ 
 from the Higgs decay and the invariant mass of $c\,\tau^+\tau^-$
from the top quark decay are used, which produce peaks near $m_{\phi}$ and $m_t$, and
(b) for $m_{\phi} > 150$ GeV, we only apply the $M_{\tau^+\tau^-}$ invariant mass 
reconstruction.

\begin{figure}[htb]
 \centering
 \includegraphics[width=68mm]{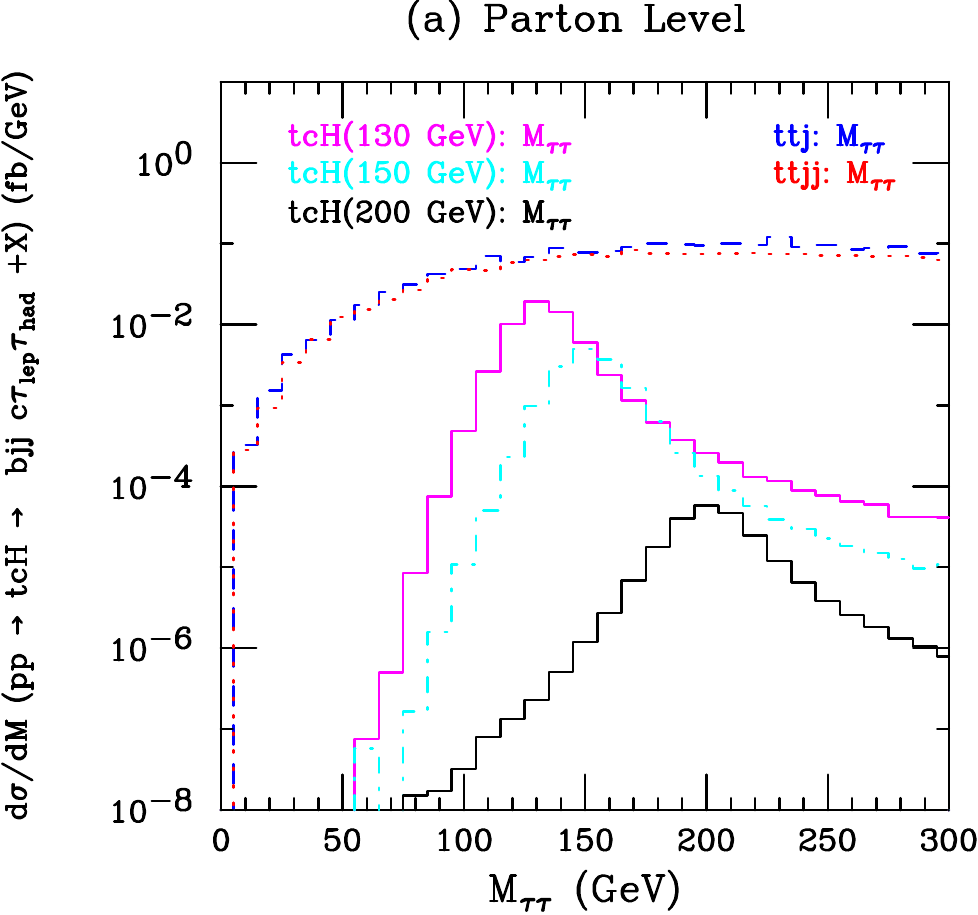}
   \hspace{1mm}
 \includegraphics[width=68mm]{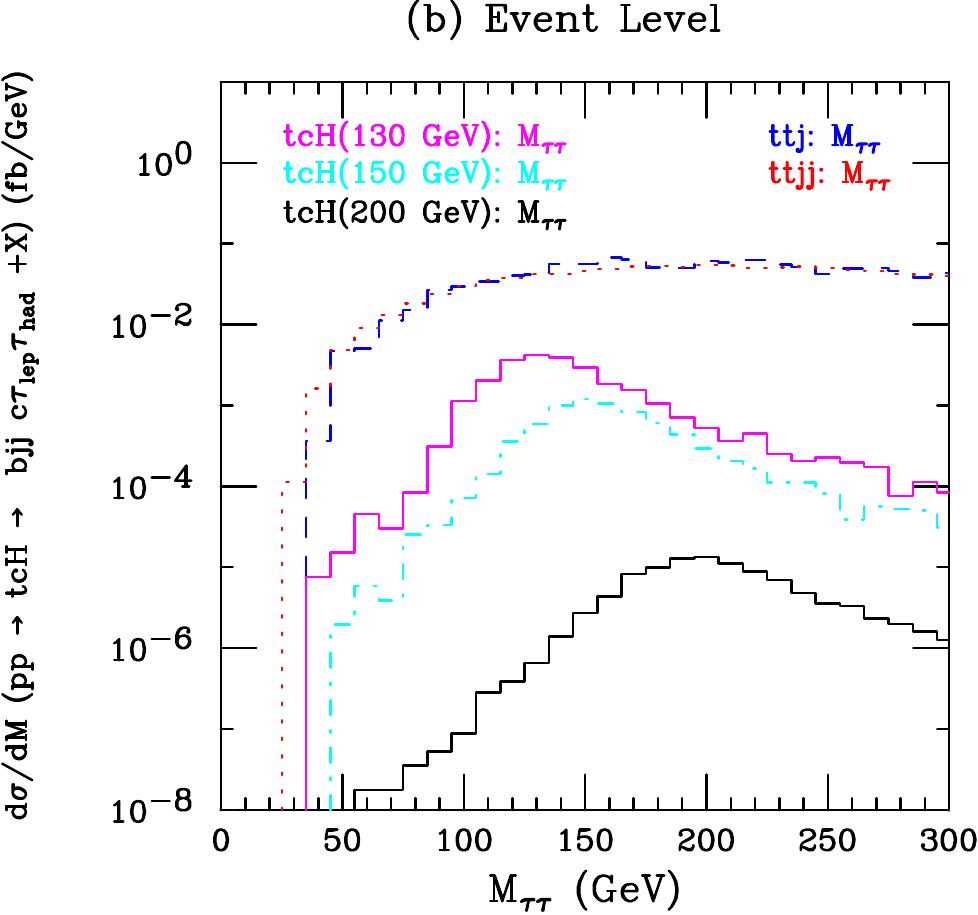}
 \caption{Invariant mass distributions $d\sigma/dM_{\tau\tau}$
     for the Higgs signal with $m_H=$ 130 GeV (magenta, solid), 150 GeV (cyan, dot-dashed), 200 GeV (black, solid) and for the dominant background from $ttj$ (blue, dashed) and $ttjj$ (red, dotted) at $\sqrt{s} = 13$ TeV.
Results are shown for (a) parton level, and (b) event level with detector simulation in $pp$ collisions.
     }
   \label{mtata}
\end{figure}

\begin{figure}[htb]
 \centering
 \includegraphics[width=68mm]{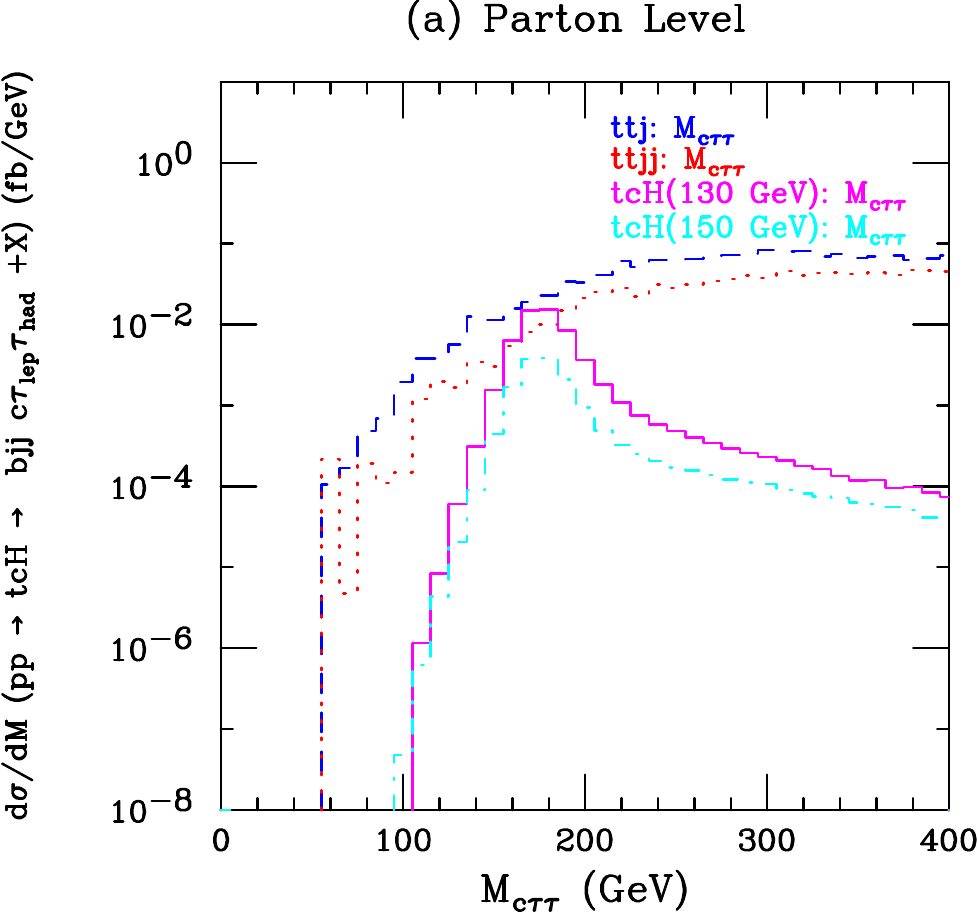}
   \hspace{1mm}
 \includegraphics[width=68mm]{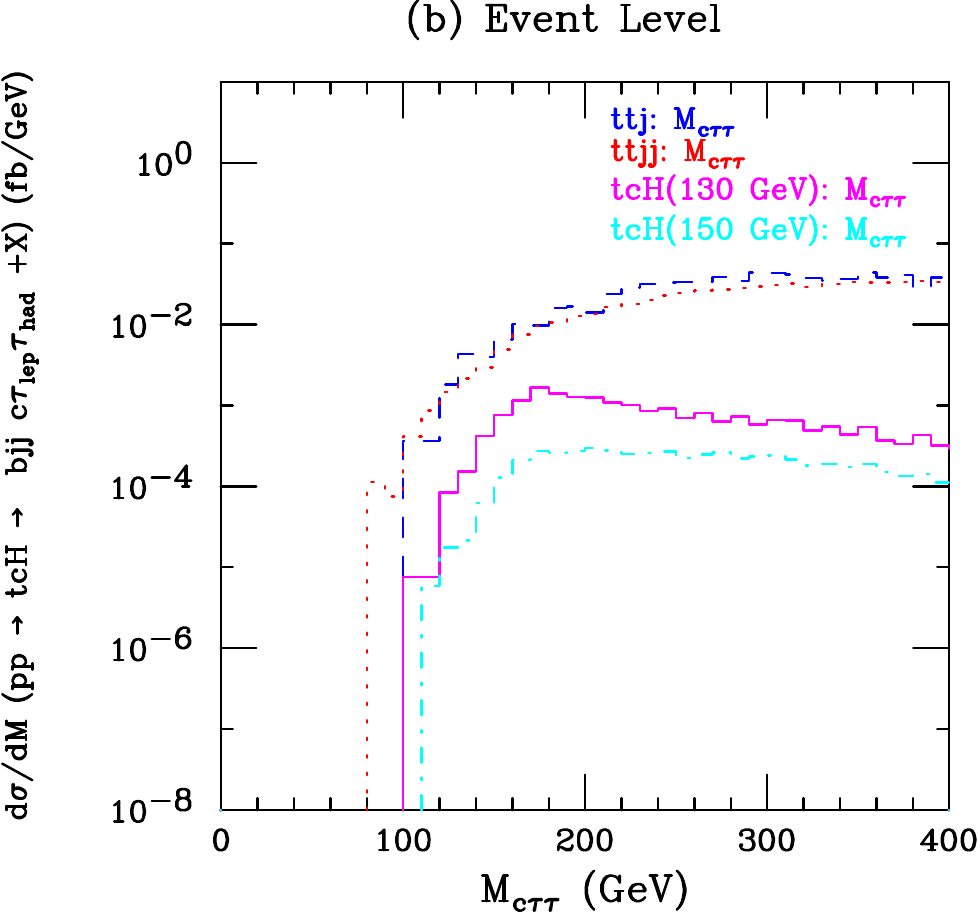}
 \caption{Invariant mass distributions $d\sigma/dM_{c\tau\tau}$
   for the Higgs signal ($t\to c\phi^0$)
   with $m_H =$ 130 GeV (magenta, solid) and 150 GeV (cyan,
   dot-dashed) and
   for the dominant background from $ttj$ (blue, dashed) and $ttjj$
   (red, dotted) at $\sqrt{s} = 13$ TeV.
   Results are shown for (a) parton level, and
     (b) event level with detector simulation in $pp$ collisions.
     }
   \label{mctata}
\end{figure}

Figure~\ref{mtata} presents the invariant mass distributions
$d\sigma/dM_{\tau \tau}$ for $m_H = 130$ GeV, 150 GeV, and 200 GeV.
There are pronounced peaks near the Higgs boson mass for all
values of the Higgs boson mass. The cross sections are evaluated for
$H^0$ with $\tilde{\rho}_{tc} = 0.1$. 
In the alignment limit, the cross section of $A^0$ is comparable to that of $H^0$.
In addition, we present the invariant mass distributions
$d\sigma/dM_{c \tau \tau}$ in Fig.~\ref{mctata}
for $m_{\phi} = 130$ GeV and 150 GeV.
There are pronounced peaks near the Higgs boson mass for $m_{\phi} \leq 150$ GeV.
However, the peak of $M_{c\tau\tau}$ becomes very broad for $m_{\phi} > 150$ GeV.
We require the reconstructed Higgs boson mass and top quark mass to
lie in the mass windows 
(a) $\Delta M_{\tau\tau} = |M_{\tau\tau}-m_{\phi}| = 0.20 m_{\phi}$ and
(b) $\Delta M_{c\tau\tau} = |M_{c\tau\tau}-m_t| = 0.25 m_t$
for $m_{\phi} < m_t - m_c$, using
the ATLAS mass resolution~\cite{Aaboud:2018pen}.

\section{The Physics Background}

The dominant background to the signal is $t\bar{t}j$ ($j = q$ or $g$), with one top quark decaying leptonically ($t \to b \ell \nu$) the other decaying hadronically ($t \to b j j$). The second most dominant background (about one third of the $t\bar{t}j$ background) is $t\bar{t}jj(+\tau),\ j = q$ or $g$ with one top quark decaying leptonically ($t \to b \ell \nu$)
and the other decaying into $\tau_{had}$ ($t \to b \tau_{had} \nu_\tau$).
These two backgrounds contribute more than 95\% of the total background.

\begin{figure}[htb]
 \centering

 \includegraphics[width=68mm]{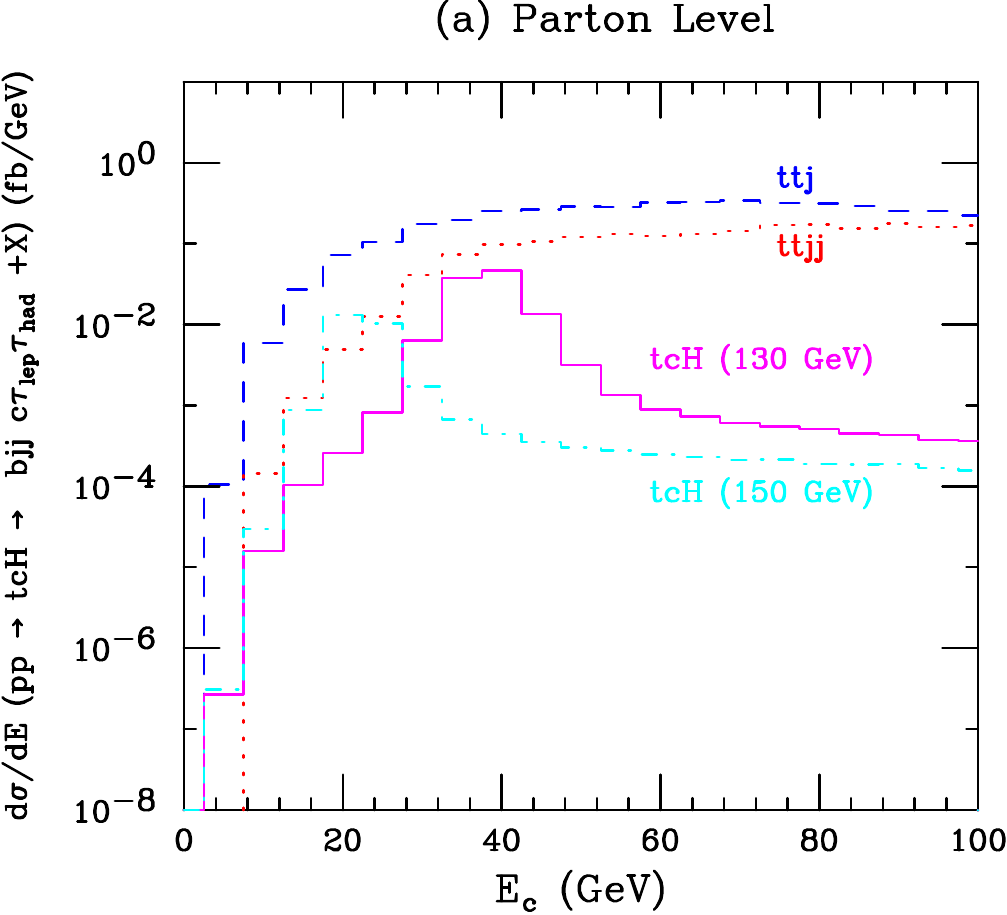}
 \hspace{1mm}
 \includegraphics[width=68mm]{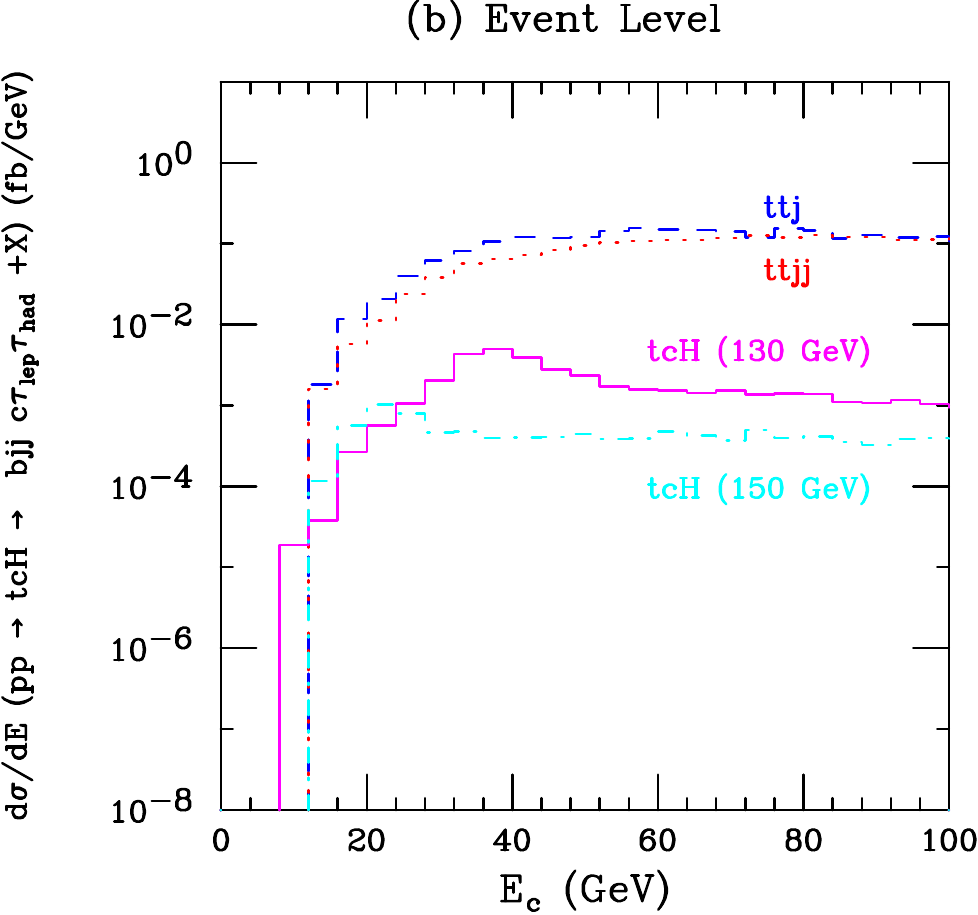}
 
 \caption{Energy distribution for the charm quark ($d\sigma/dE_c$)
   in the top quark rest frame for the Higgs signal
   from $t\to c H^0$ with $m_H=$ 130 GeV (magenta, solid),
   150 GeV (cyan, dot-dashed) and for the dominant background from
   $ttj$ (blue, dashed) and $ttjj$ (red, dotted)
   at $\sqrt{s} = 13$ TeV.
   Results are shown for (a) parton level, and
   (b) event level with detector simulation in $pp$ collisions.
   }
 \label{fig:Echarm}
\end{figure}

In addition to all of the kinematic requirements and the invariant masses, 
the energy of the charm quark ($E^*_c$) in the rest frame of the top quark
can be reconstructed to discriminate the  $t \to c \phi^0$
signal from the background~\cite{Han:2001ap,Kao:2011aa}.
For $m_{\phi} < m_t$, we can calculate the $E_c^*$ from the following equation:
\begin{equation}
    E_c^* = \frac{m_t}{2}\left[1 + \frac{m_c^2}{m_t^2} 
           -\frac{m_{\phi}^2}{m_t^2}  \right] \, .
\end{equation}
Figure~\ref{fig:Echarm} presents the energy distributions of the
charm quark in the top quark rest frame for the two main backgrounds
and for $m_\phi = 130$ GeV and 150 GeV,
which exhibit enhancements at 37.4~GeV and 21.2~GeV, respectively.
Selecting events with $E^*_c < 60$~GeV removes a larger
fraction of background events than signal events, thereby enhancing
the discovery potential.


From the invariant mass and the charm quark energy distributions  
at the parton and event levels, the following mass requirements
are deduced
\begin{itemize}
\item[(i)] $|M_{j_1 j_2} - m_W| \leq 0.20 \times m_W$ and
  $|M_{b j_1 j_2} - m_t | \leq 0.25 \times m_t$,
\item[(ii)] $|M_{\tau\tau} - m_{\phi}| \leq 0.20 \times m_{\phi}$,
\item[(iii)] $|M_{c\tau\tau} - m_t| \leq 0.25 \times m_t$
  for $m_{\phi} \leq 150$ GeV,
\item[(iv)] 0 GeV $\leq E^*_c \leq $ 60 GeV for $m_{\phi} \leq 150$ GeV.
\end{itemize}
For the case $m_{\phi}>150$ GeV, we only apply (i) and (ii)
to the signal and background events.
These requirements are chosen to reduce the physics background
in a manner that maximizes the statistical significance of the signal.

\section{Discovery Potential at the Parton Level and Event Level}

%
%
\begin{table}[htb]
 \centering
 \begin{tabular}{ccccc} \hline 
 Selection Criteria  & $\text{S}_1 \,\,\,$  & $\text{BG}_1 \,\,\,$& $\text{S}_2 \,\,\,$ & $\text{BG}_2$ \\
   \hline 
   $\ge$ 5 jets, 1 lepton & \checkmark & \checkmark & \checkmark & \checkmark \\
   1 tagged $b$-jet, 1 tagged $\tau$-jet & \checkmark & \checkmark & \checkmark & \checkmark \\
   $p_T(b,j)>25$ GeV, $|\eta(j)|<2.5$ & \checkmark & \checkmark & \checkmark & \checkmark \\
   $p_T(l)>20$ GeV, $|\eta(l)|<2.5$ & \checkmark & \checkmark & \checkmark & \checkmark \\
   $\Delta R>$0.4 & \checkmark & \checkmark & \checkmark & \checkmark \\
   $\slashed{E}_T>30$ GeV & \checkmark & \checkmark & \checkmark & \checkmark \\
   $|M_{j_1 j_2} - m_W| \leq 0.20 \times m_W$ & \checkmark & \checkmark & \checkmark & \checkmark \\
   $|M_{b j_1 j_2} - m_t | \leq 0.25 \times m_t$ & \checkmark & \checkmark & \checkmark & \checkmark \\
   $|M_{\tau\tau} - m_{\phi}| \leq 0.20 \times m_{\phi}$ & \checkmark & \checkmark & \checkmark & \checkmark \\
   $|M_{c\tau\tau} - m_t| \leq 0.25 \times m_t$ & \checkmark & \checkmark & $-$ & $-$ \\
   0 GeV $\leq E^*_c \leq $ 60 GeV & \checkmark & \checkmark & $-$ & $-$ \\
 \hline 
 \end{tabular}
 \caption{Selection criteria for signal and backgrounds events for $m_\phi\le150$ GeV and $m_\phi>150$ GeV. In this table, $\text{S}_1$ represents signal events with $m_\phi\le150$~GeV and $\text{BG}_1$ its background, $\text{S}_2$ represents signal events with $m_\phi>150$~GeV and $\text{BG}_2$ its background.}
 \label{allcuts}
\end{table}

We present the parton level signal cross sections at 
$\sqrt{s} = 13$, 13.6, and 14 TeV, after
applying all the selection criteria (see Table~\ref{allcuts}) for $\tilde{\rho}_{tc} = 0.1$ and $0.5$,
in Table~\ref{psigcross}.
The cross sections for the backgrounds
after applying the selection requirements
are presented in Table~\ref{pbkgcross}.

%
%
\begin{table}[htb]
 \centering
 \begin{tabular}{ccc} \hline \hline
 $\sqrt{s}=13$ TeV  & $\quad\rho_{tc} = 0.1 \quad$ & $\rho_{tc} = 0.5$ \\
   \hline \hline  
 $m_{H}=$ 130 GeV $\quad$ & 0.3857 fb $\quad$ & 9.657 fb \\
 $m_{H}=$ 150 GeV $\quad$ & 0.1166 fb $\quad$ & 2.889 fb \\
 $m_{H}=$ 200 GeV $\quad$ & 0.0016 fb $\quad$ & 0.0023 fb \\
 \hline \hline
 $\sqrt{s}=13.6$ TeV & $\quad\rho_{tc} = 0.1 \quad$ & $\rho_{tc} = 0.5$ \\
   \hline \hline  
 $m_{H}=$ 130 GeV $\quad$ & 0.4241 fb $\quad$ & 10.58 fb \\
 $m_{H}=$ 150 GeV $\quad$ & 0.1300 fb $\quad$ & 3.231 fb \\
 $m_{H}=$ 200 GeV $\quad$ & 0.0018 fb $\quad$ & 0.0025 fb \\
 \hline \hline
 $\sqrt{s}=14$ TeV & $\quad\rho_{tc} = 0.1 \quad $ & $\rho_{tc} = 0.5$ \\
   \hline \hline  
 $m_{H}=$ 130 GeV $\quad$ & 0.4567 fb $\quad$ & 11.47 fb \\
 $m_{H}=$ 150 GeV $\quad$ & 0.1370 fb $\quad$ & 3.434 fb \\
 $m_{H}=$ 200 GeV $\quad$ & 0.0020 fb $\quad$ & 0.0027 fb \\
 \hline \hline
 \end{tabular}
 \caption{Parton level signal cross sections
   after all selection criteria are applied
   and $b$-jet and tau-jet efficiencies $\varepsilon_b = 0.7$ and
   $\varepsilon_\tau = 0.723$, respectively.}
 \label{psigcross}
\end{table}

%
%
\begin{table}[htb]
  \centering
  \begin{tabular}{ccccc}
  \hline \hline
  $\sqrt{s}=13$ TeV & $t \bar{t} j$ & $t \bar{t}jj (+\tau)$ & $Zb\bar{b}jj$ & Total  \\
  \hline\hline
  $m_{H}=130$ GeV $\quad$ & 0.7236 fb$\quad$ & 0.1458 fb$\quad$ & 5.4$\times 10^{-2}$ fb $\quad$ & 0.8753 fb  \\
  $m_{H}=150$ GeV $\quad$ & 4.826 fb $\quad$ & 1.144 fb $\quad$ & 3.6$\times 10^{-2}$ fb $\quad$ & 6.005 fb  \\
  $m_{H}=200$ GeV $\quad$ & 8.145 fb $\quad$ & 1.668 fb $\quad$ & 1.2$\times 10^{-2}$ fb $\quad$ & 9.824 fb  \\ \hline \hline
  $\sqrt{s}=13.6$ TeV & $t \bar{t} j$ & $t \bar{t}jj (+\tau)$ & $Zb\bar{b}jj$ & Total  \\
  \hline\hline
  $m_{H}=130$ GeV $\quad$ & 0.8010 fb$\quad$ & 0.1568 fb$\quad$ & 5.9$\times 10^{-2}$ fb$\quad$ & 0.9640 fb \\
  $m_{H}=150$ GeV $\quad$ & 5.326 fb $\quad$ & 1.281 fb $\quad$ & 4.0$\times 10^{-2}$ fb$\quad$ & 6.647 fb  \\
  $m_{H}=200$ GeV $\quad$ & 9.052 fb $\quad$ & 1.849 fb $\quad$ & 1.1$\times 10^{-2}$ fb$\quad$ & 10.91 fb \\ 
  \hline \hline
  $\sqrt{s}=14$ TeV & $t \bar{t} j$ & $t \bar{t}jj (+\tau)$ & $Zb\bar{b}jj$ & Total  \\
  \hline \hline
  $m_{H}=130$ GeV $\quad$ & 0.8586 fb$\quad$ & 0.1667 fb$\quad$ & 6.3$\times 10^{-2}$ fb$\quad$& 1.031 fb  \\
  $m_{H}=150$ GeV $\quad$ & 5.683 fb $\quad$ & 1.378 fb $\quad$ & 4.3$\times 10^{-2}$ fb$\quad$& 7.103 fb \\
  $m_{H}=200$ GeV $\quad$ & 9.639 fb $\quad$ & 1.978 fb $\quad$ & 1.2$\times 10^{-2}$ fb$\quad$& 11.63 fb  \\ \hline \hline
  \end{tabular}
  \caption{Background cross sections after applying the mass selection
    at the parton level. }
  \label{pbkgcross}
\end{table}

Applying all the selection requirements, we present the signal cross
sections at the event level in Table~\ref{esigcross} for
$m_{H} = 130$ GeV, 150 GeV, and 200 GeV as well as $\tilde{\rho}_{tc} = 0.1$ and
0.5.
The event level cross sections for the dominant physics backgrounds
are shown in Table~\ref{ebkgcross}.
We evaluate the statistical significance ($N_{SS}$) as a function of
$\tilde{\rho}_{tc}$ and $m_{\phi}$ from this study, where
$N_{SS}$ is calculated using \cite{Cowan:2010js}
\begin{equation}
    N_{SS} = \sqrt{ 2 (N_S + N_B)\ln(1 + N_S/N_B) - 2 N_S }.
\end{equation}  
Here $N_{S}$ and $N_{B}$ are number of signal and
background events, where $N_{S}= \sigma_s\times L$, $N_{B}=
\sigma_b\times L$, and $L$ is the integrated luminosity of the LHC.
%
%
\begin{table}[htb]
 \centering
 \begin{tabular}{ccc} \hline \hline
 $\sqrt{s}=13$ TeV  & $\quad\rho_{tc} = 0.1 \quad$ & $\rho_{tc} = 0.5$ \\
   \hline \hline  
 $m_{H}=$ 130 GeV $\quad$ & 0.1589 fb $\quad$ & 3.474 fb \\
 $m_{H}=$ 150 GeV $\quad$ & 0.0907 fb $\quad$ & 2.317 fb \\
 $m_{H}=$ 200 GeV $\quad$ & 1.1$\times 10^{-3}$ fb $\quad$ & 1.5$\times 10^{-3}$ fb \\
 \hline \hline
 $\sqrt{s}=13.6$ TeV & $\quad\rho_{tc} = 0.1 \quad$ & $\rho_{tc} = 0.5$ \\
   \hline \hline  
 $m_{H}=$ 130 GeV $\quad$ & 0.1685 fb $\quad$ & 4.108 fb \\
 $m_{H}=$ 150 GeV $\quad$ & 0.1026 fb $\quad$ & 2.677 fb \\
 $m_{H}=$ 200 GeV $\quad$ & 1.2$\times 10^{-3}$ fb $\quad$ & 1.8$\times 10^{-3}$ fb \\
 \hline \hline
 $\sqrt{s}=14$ TeV & $\quad\rho_{tc} = 0.1 \quad $ & $\rho_{tc} = 0.5$ \\
   \hline \hline  
 $m_{H}=$ 130 GeV $\quad$ & 0.1894 fb $\quad$ & 8.050 fb \\
 $m_{H}=$ 150 GeV $\quad$ & 0.1031 fb $\quad$ & 5.638 fb \\
 $m_{H}=$ 200 GeV $\quad$ & 1.4$\times 10^{-3}$ fb $\quad$ & 2.0$\times 10^{-3}$ fb \\
 \hline \hline
 \end{tabular}
 \caption{ Event level signal cross sections after all selections are applied.}
 \label{esigcross}
\end{table}

%
%
\begin{table}[htb]
  \centering
  \begin{tabular}{ccccc}
  \hline \hline
  $\sqrt{s}=13$ TeV & $t \bar{t} j$ & $t \bar{t}jj (+\tau)$ & $Zb\bar{b}jj$ & Total  \\
  \hline\hline
  $m_{H}=130$ GeV $\quad$ & 0.7157 fb$\quad$ & 0.3559 fb$\quad$ & 1.6$\times 10^{-2}$ fb$\quad$ & 1.088 fb  \\
  $m_{H}=150$ GeV $\quad$ & 4.689 fb $\quad$ & 1.904 fb $\quad$ & 4.1$\times 10^{-2}$ fb$\quad$ & 6.635 fb  \\
  $m_{H}=200$ GeV $\quad$ & 6.860 fb $\quad$ & 2.720 fb $\quad$ & 1.4$\times 10^{-2}$ fb$\quad$ & 9.594 fb  \\ \hline \hline
  $\sqrt{s}=13.6$ TeV & $t \bar{t} j$ & $t \bar{t}jj (+\tau)$ & $Zb\bar{b}jj$ & Total  \\
  \hline\hline
  $m_{H}=130$ GeV $\quad$ & 0.8734 fb$\quad$ & 0.3982 fb$\quad$ & 2.5$\times 10^{-2}$ fb$\quad$ & 1.297 fb  \\
  $m_{H}=150$ GeV $\quad$ & 5.098 fb $\quad$ & 2.173 fb $\quad$ & 4.3$\times 10^{-2}$ fb$\quad$ & 7.313 fb  \\
  $m_{H}=200$ GeV $\quad$ & 7.065 fb $\quad$ & 3.200 fb $\quad$ & 2.2$\times 10^{-2}$ fb$\quad$ & 10.29 fb  \\ 
  \hline \hline
  $\sqrt{s}=14$ TeV & $t \bar{t} j$ & $t \bar{t}jj (+\tau)$ & $Zb\bar{b}jj$ & Total  \\
  \hline \hline
  $m_{H}=130$ GeV $\quad$ & 0.9331 fb$\quad$ & 0.4034 fb$\quad$ & 2.7$\times 10^{-2}$ fb$\quad$ & 1.364 fb  \\
  $m_{H}=150$ GeV $\quad$ & 5.544 fb $\quad$ & 2.266 fb $\quad$ & 5.0$\times 10^{-2}$ fb$\quad$ & 7.859 fb  \\
  $m_{H}=200$ GeV $\quad$ & 7.367 fb $\quad$ & 3.204 fb $\quad$ & 2.9$\times 10^{-2}$ fb$\quad$ & 10.60 fb  \\ \hline \hline
  \end{tabular}
  \caption{Background cross sections after applying the mass selection at the event level.}
  \label{ebkgcross}
\end{table}

\begin{figure}[htb]
 \centering
 \includegraphics[width=68mm]{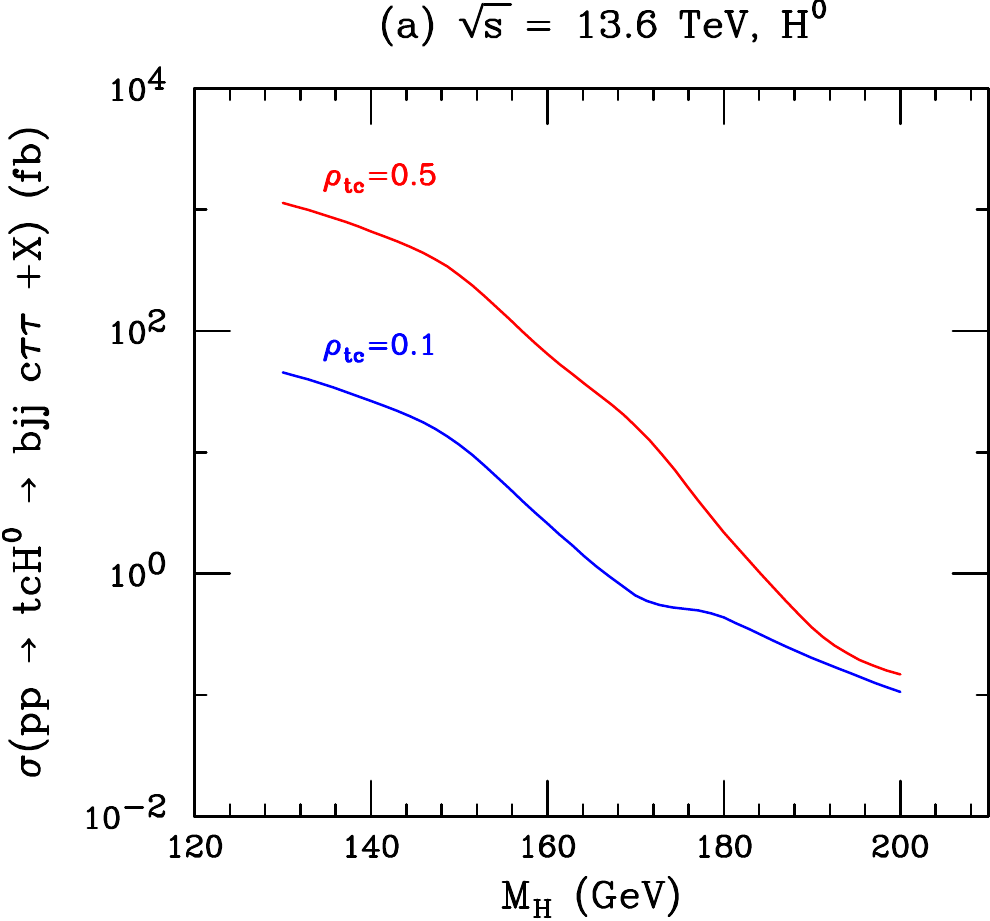}
   \hspace{1mm}
 \includegraphics[width=68mm]{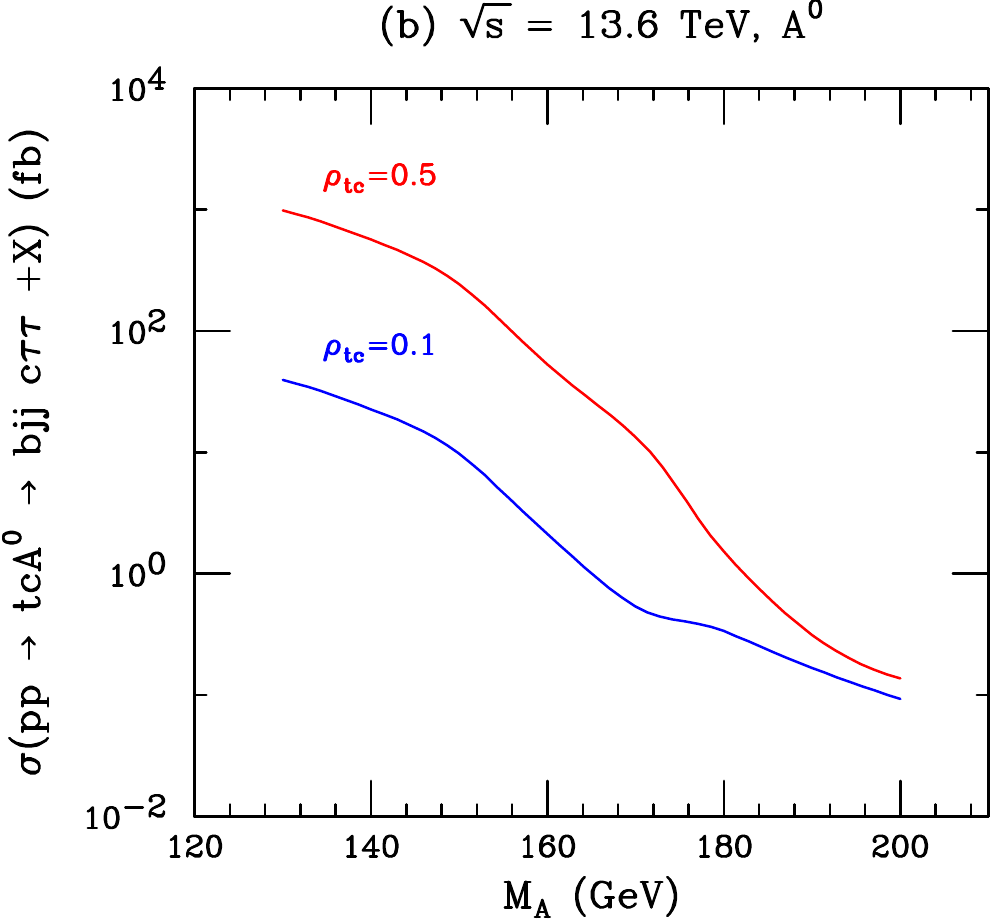}
 \caption{Signal cross section $\sigma$ vs. mass of Higgs boson for the FCNH signal ($t\to c\Phi^0$) with $\rho_{tc}=$ 0.1 (blue) and 0.5 (red) at $\sqrt{s}=13.6$ TeV, 
where $\phi^0$ is (a) Heavy Scalar $H^0$, and (b) pseudoscalar $A^0$ with detector simulation in $pp$ collisions.}
   \label{sigvsm}
\end{figure}

\begin{figure}[htb]
 \centering
 \includegraphics[width=68mm]{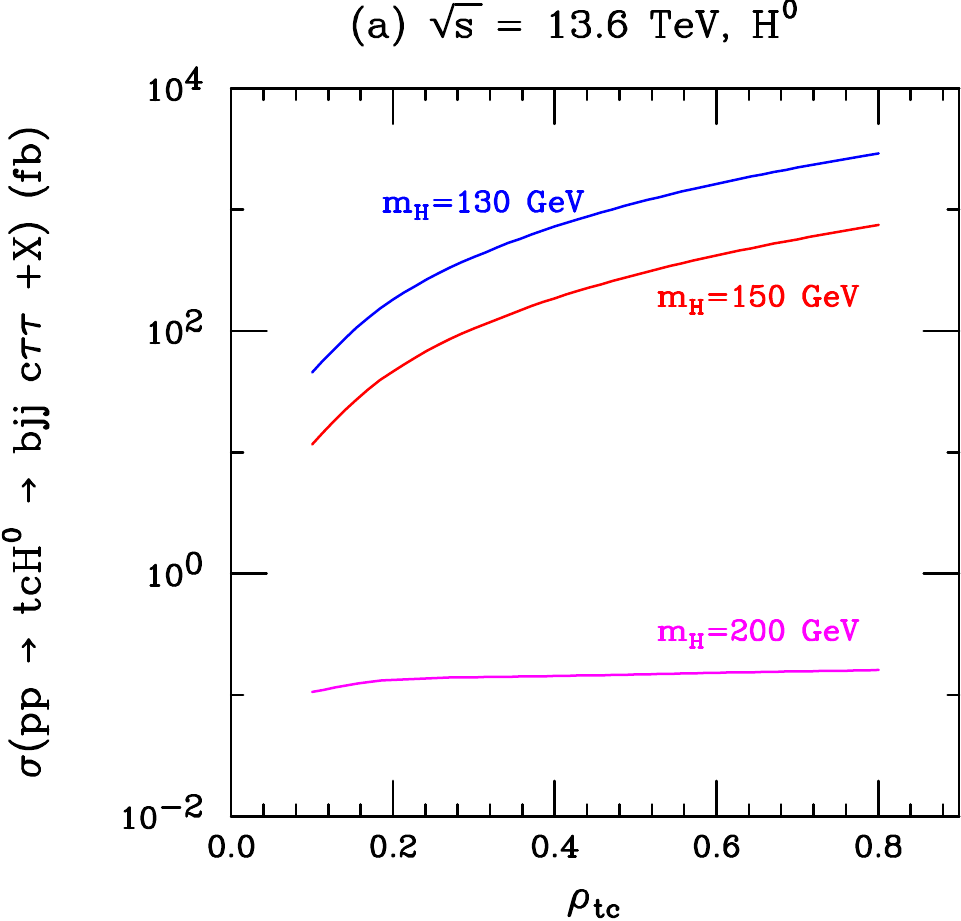}
   \hspace{1mm}
 \includegraphics[width=68mm]{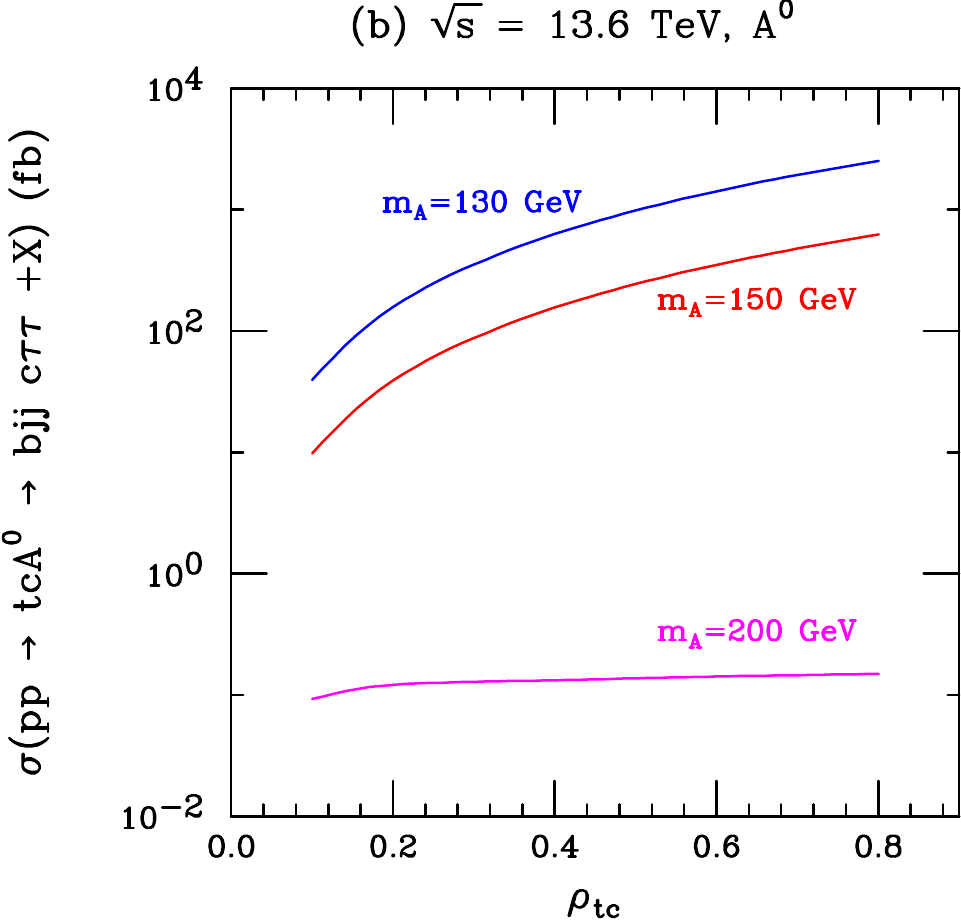}
 \caption{Signal cross section  $\sigma$ vs. $\rho_{tc}$ of Higgs boson for the FCNH signal ($t\to c\Phi^0$) with $m_\phi=$ 130 GeV (blue), 150 GeV (red) and 200 GeV (magenta) at $\sqrt{s}=13.6$ TeV, 
where $\Phi^0$ is (a) Heavy Scalar $H^0$, and (b) pseudoscalar $A^0$ with detector simulation in $pp$ collisions.}
   \label{sigvsrho}
 \end{figure}

\begin{figure}[htb]
 \centering

 \includegraphics[height=60mm]{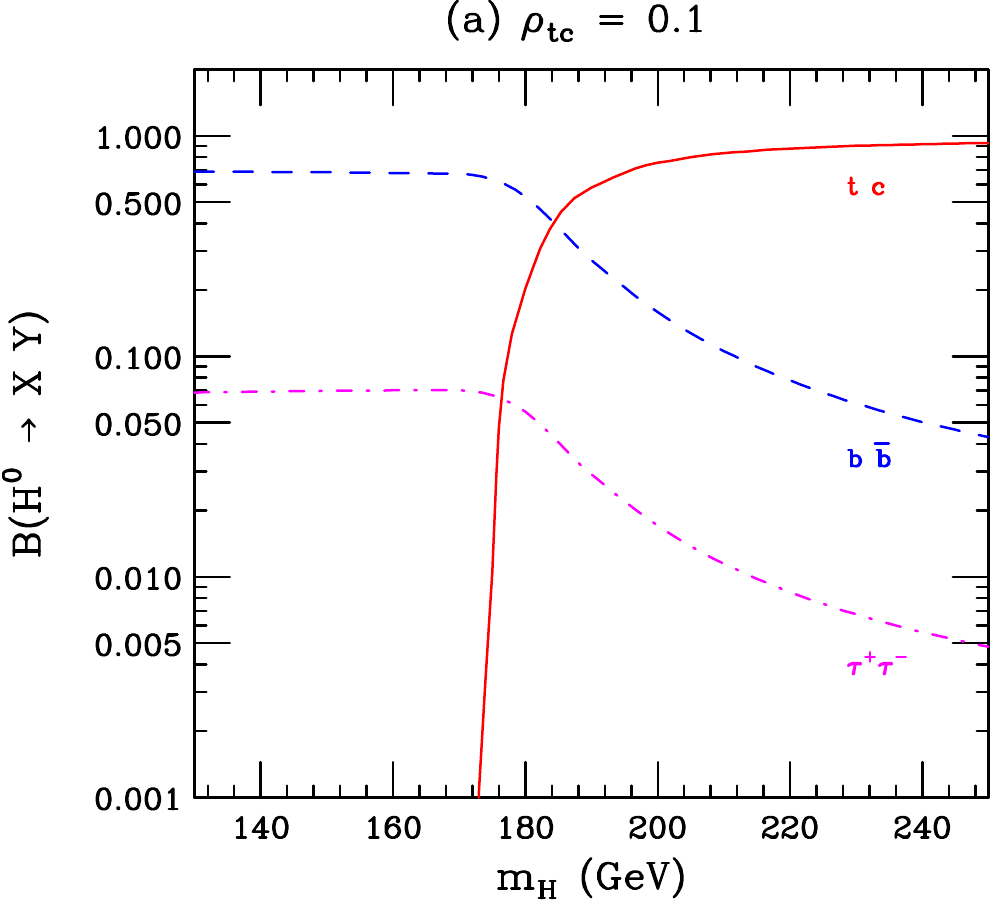}
 \includegraphics[height=60mm]{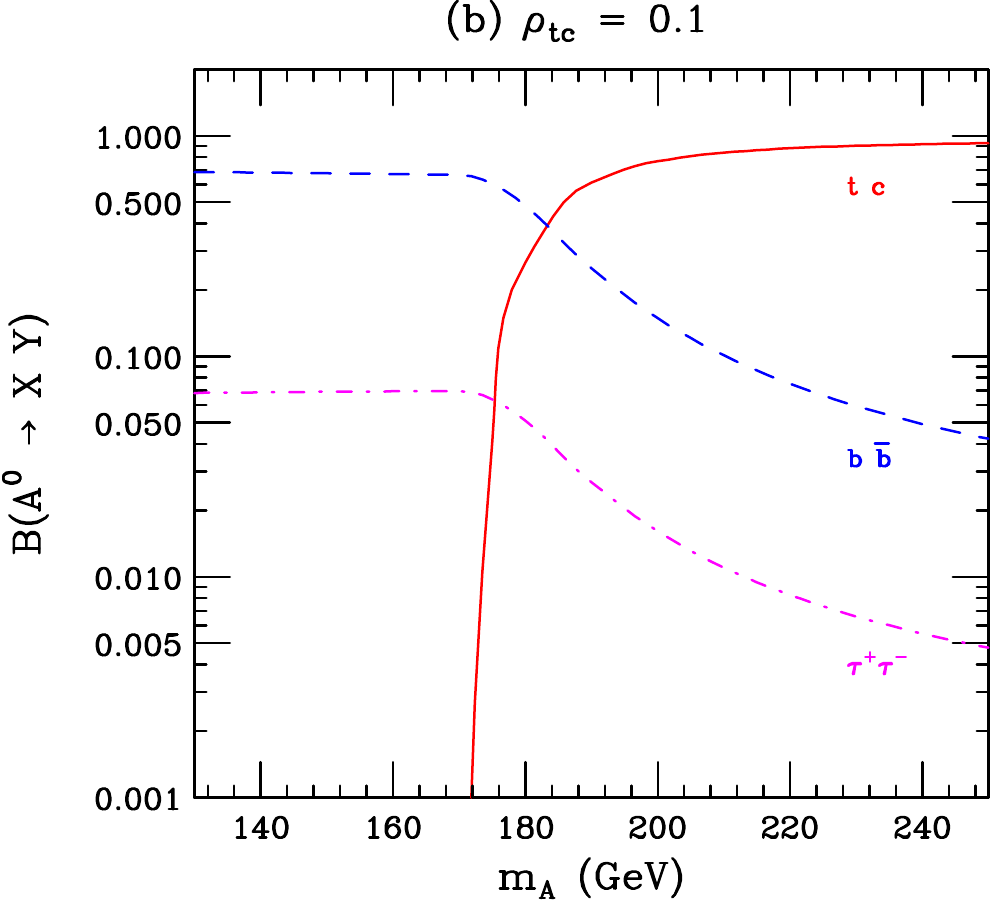}
\\
 \includegraphics[height=60mm]{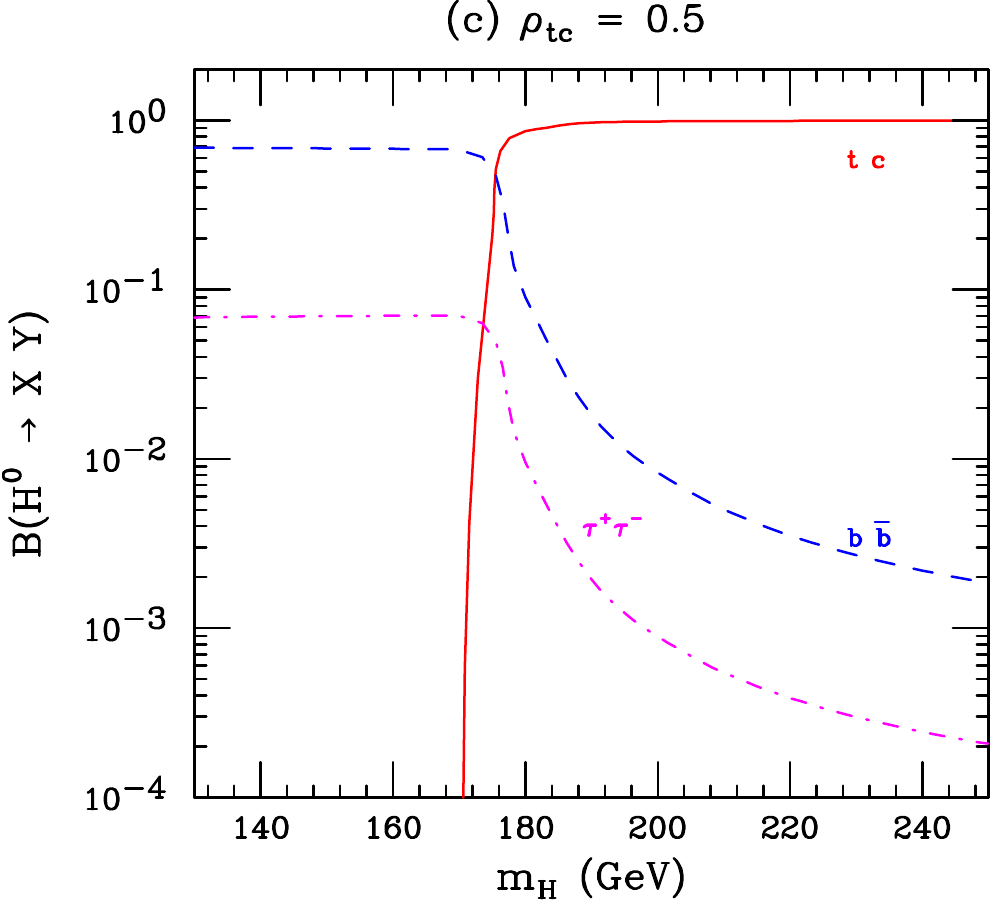}
 \includegraphics[height=60mm]{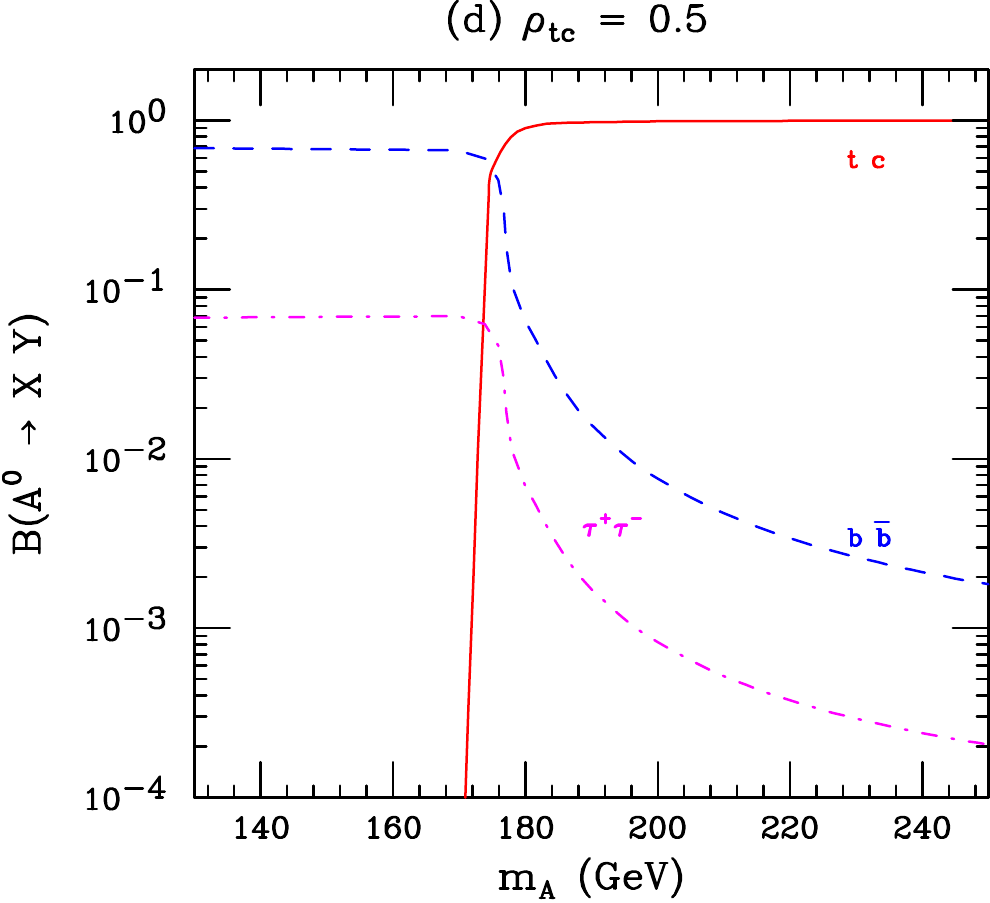}

 \caption{Branching ratios of $\phi^0$ as functions of $m_\phi$ for the following cases: (a) $H^0$ with $\tilde{\rho}_{tc} = 0.1$, (b) $A^0$ with $\tilde{\rho}_{tc} = 0.1$, (c) $H^0$ with $\tilde{\rho}_{tc} = 0.5$, and (d) $A^0$ with $\tilde{\rho}_{tc} = 0.5$.}
   \label{brh2h3}
\end{figure}

Figure~\ref{sigvsm} presents the signal cross section as a function of
the heavy Higgs boson mass
for $\tilde{\rho}_{tc} = 0.1$ and 0.5. 
In the alignment limit of 2HDM with $\sin(\beta-\alpha) \to 1$,
the cross section for $H^0$ production
is comparable to that of the $A^0$.
For the case that $m_{H} \simeq m_{A}$, the signal cross section
and the statistical significance is doubled.
We observe that the cross section declines rapidly as $m_{H}$
increases for $m_{H} < m_t$ and $m_{H}$ is closer to $m_t$ 
as the charm quark becomes less energetic,
but decreases slowly for $m_{H} > m_t$. 
This indicates that the impact of $m_{H}$ becomes less significant
when $t$ or $\bar{t}$ become virtual and
$H^0 \to t\bar{c}$ decay channel opens up.
Figure~\ref{sigvsrho} presents the signal cross section as
a function of $\tilde{\rho}_{tc}$ for $m_{H} =$ 130, 150, and 200 GeV.
For $m_{H} < m_t$,
the cross section increases approximately  as $\sqrt{\tilde{\rho}_{tc}}$.
However, for $m_{H} > m_t$, the cross section is approximately constant.
This indicates for $m_{H} > m_t$ the $H^0\to t\bar{c}$ decay channel
suppresses the $H^0 \to \tau^+\tau^-$ branching fraction.
That means the enhancement from $\tilde{\rho}_{tc}^2$ in the production of 
$pp \to t\bar{t} \to t c\phi^0 \to t c\tau^+\tau^- +X$ is almost
cancelled by the total width when $H^0 \to t\bar{c}$ becomes the dominant decay
channel for $\tilde{\rho}_{tc} \agt 0.1$.
Using 2HDMC~\cite{Eriksson:2009ws}, we calculate and present the
branching fractions of the relevant Higgs decay channels
in Fig.~\ref{brh2h3}.

\begin{figure}[htb]
    \centering
    \includegraphics[width=80mm]{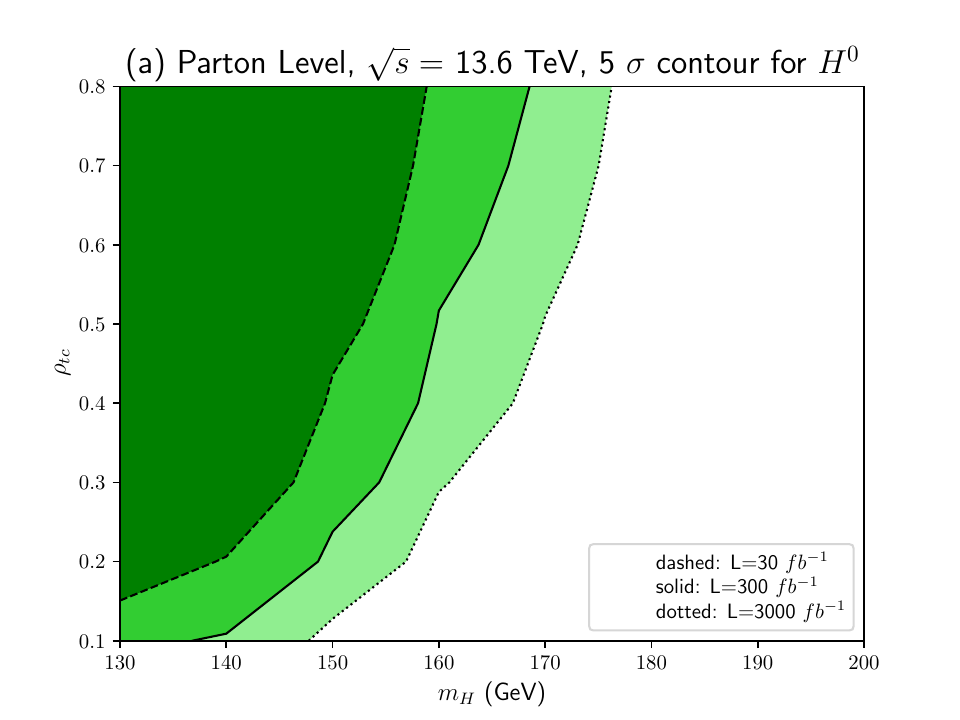}\hspace{1mm}
    \includegraphics[width=80mm]{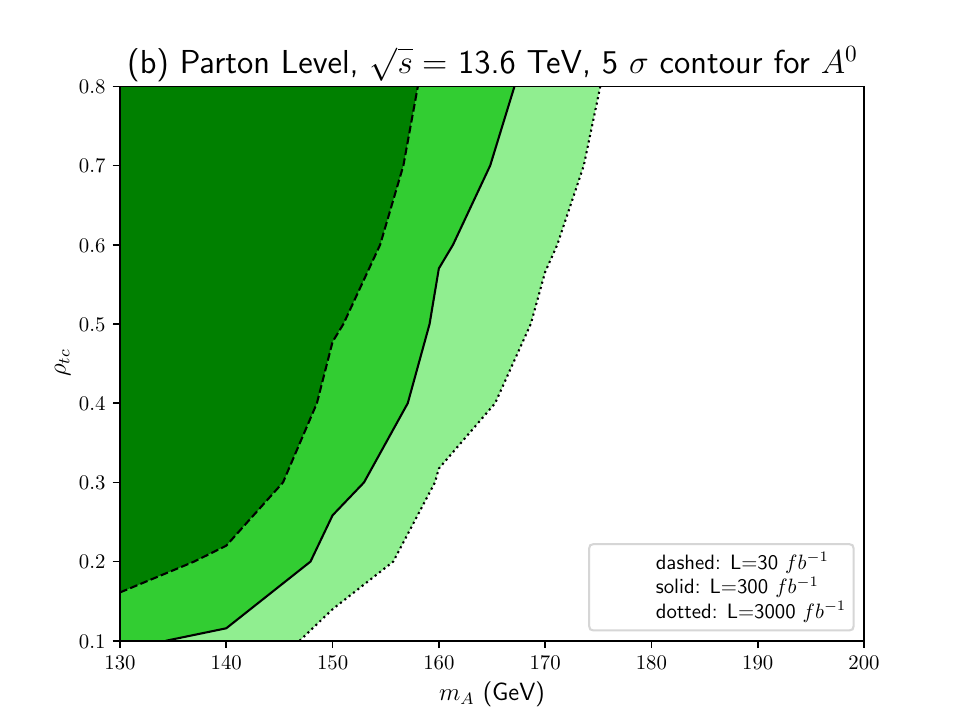}\\
    \includegraphics[width=80mm]{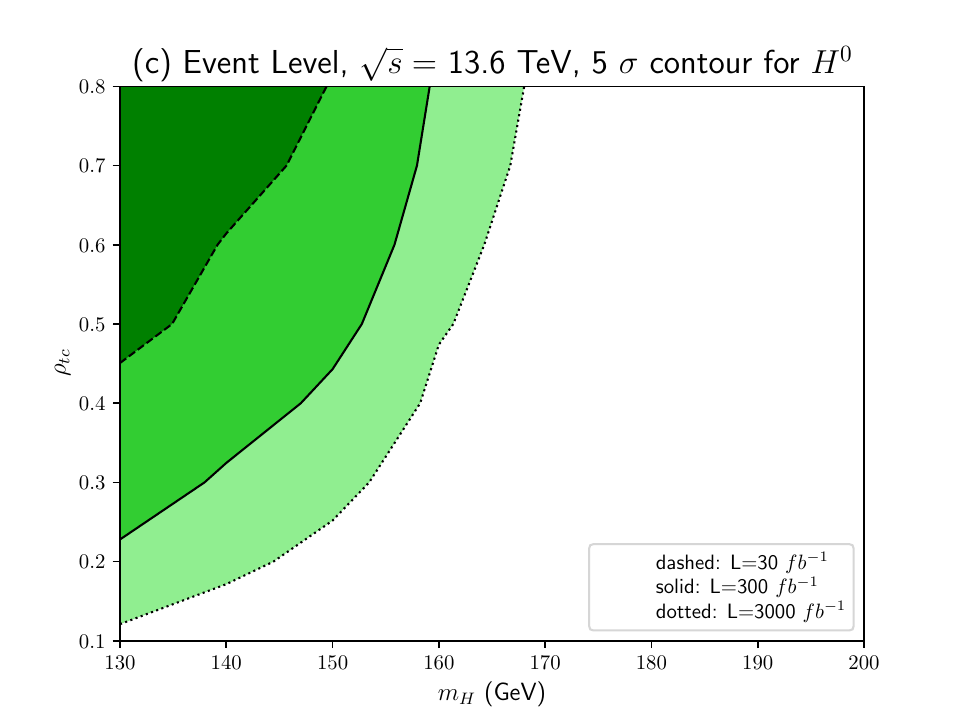}\hspace{1mm}
    \includegraphics[width=80mm]{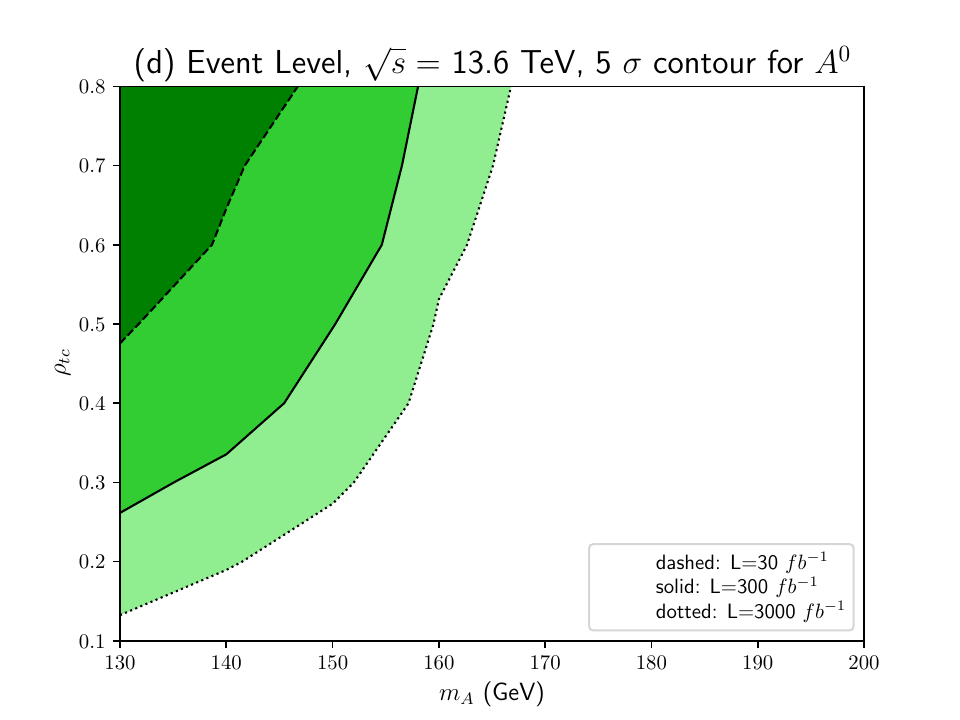}\\

    \caption{Parton Level and Event Level 5$\sigma$ discovery contours at the LHC at $\sqrt{s}=13.6$ TeV in the $[m_\Phi,\tilde{\rho}_{tc}]$ plane
      for (a) Parton Level Heavy Scalar $H^0$, (b) Parton Level pseudoscalar $A^0$, (c) Event Level Heavy Scalar $H^0$ and (d) Event Level pseudoscalar $A^0$ with $L = 30 fb^{-1}$ (dark green dotted dashed),
      $300 fb^{-1}$ (medium green solid) and
      $L = 3000 fb^{-1}$ (light green dashed)}
    \label{fig:5contour136}
\end{figure}

\begin{figure}[htb]
    \centering
    \includegraphics[width=80mm]{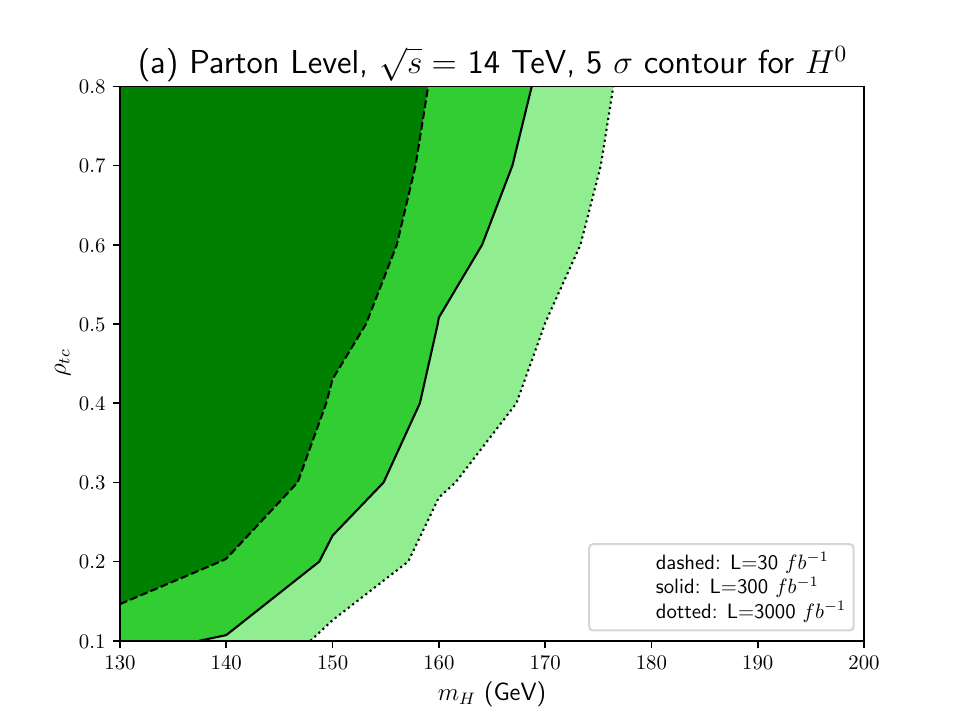}\hspace{1mm}
    \includegraphics[width=80mm]{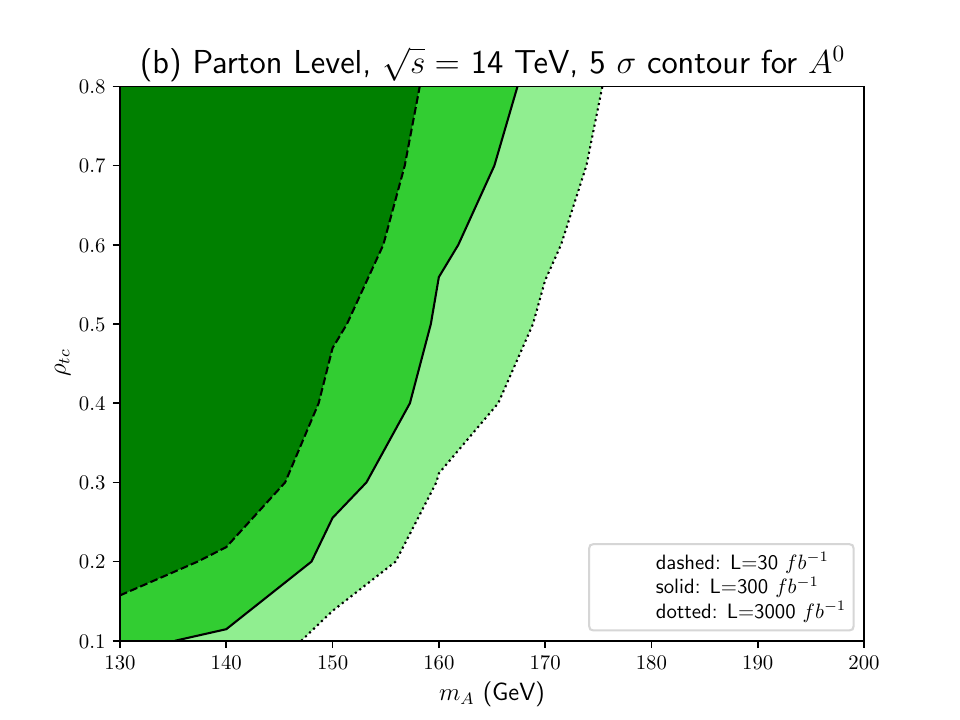}\\
    \includegraphics[width=80mm]{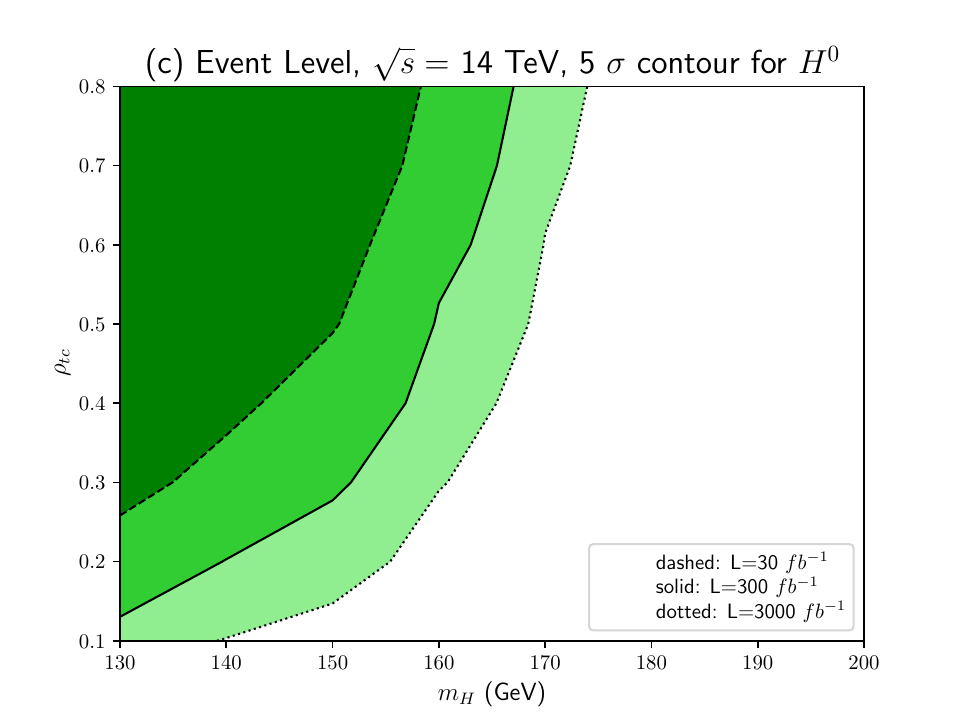}\hspace{1mm}
    \includegraphics[width=80mm]{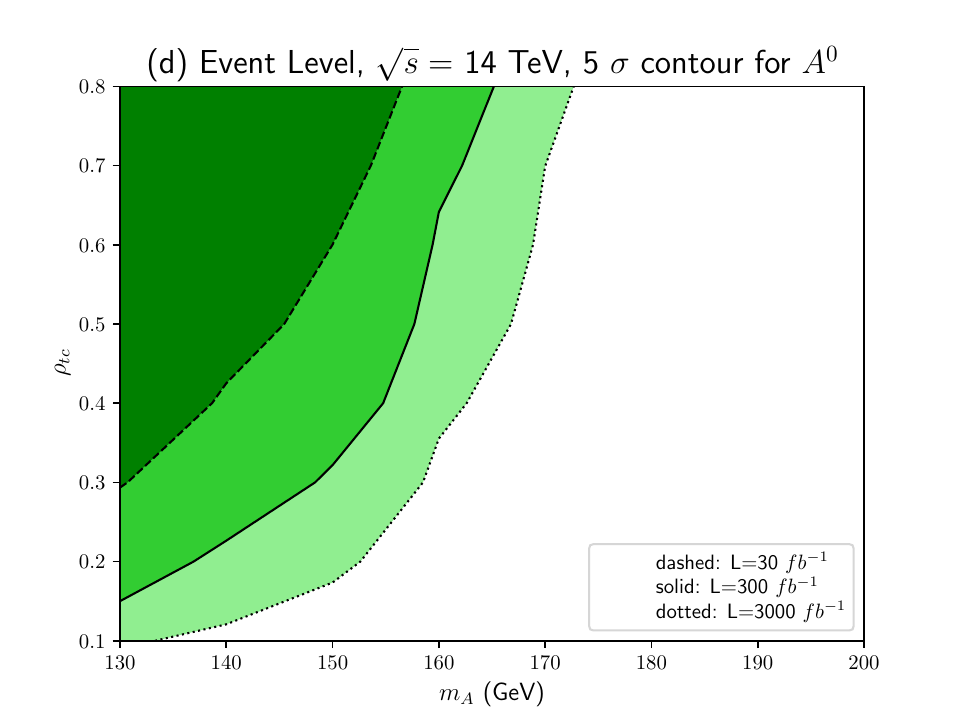}\\
    \caption{Parton Level and Event Level 5$\sigma$ discovery contours at the LHC at $\sqrt{s}=14$ TeV in the $[m_\Phi,\tilde{\rho}_{tc}]$ plane
      for (a) Parton Level Heavy Scalar $H^0$, (b) Parton Level pseudoscalar $A^0$, (c) Event Level Heavy Scalar $H^0$ and (d) Event Level pseudoscalar $A^0$ with $L = 30 fb^{-1}$ (dark green dotted dashed),
      $300 fb^{-1}$ (medium green solid) and
      $L = 3000 fb^{-1}$ (light green dashed)}
    \label{fig:5contour14}
\end{figure}

Figure~\ref{fig:5contour136} presents the $5\sigma$ discovery
significance at the LHC for $\sqrt{s} =$ 13.6 TeV and
Fig.~\ref{fig:5contour14} presents the discovery contours
for $\sqrt{s} =$ 14 TeV. 
The discovery potential are  shown for both parton level and event level in the
$[m_{\phi},\tilde{\rho}_{tc}]$ plane.
We have chosen three values of the integrated luminosity
$\mathcal{L} =$ 30, 300 and 3000~fb$^{-1}$.
It is clear for masses of Higgs boson less than the mass of top quark,
the high luminosity LHC at $\sqrt{s} = 14$ TeV
with an integrated luminosity $L = 3000 \; {\rm fb}^{-1}$ significantly
improves the discovery potential for the FCNH signal.


\section{Conclusions}

In a general 2HDM~\cite{Davidson:2005cw,Mahmoudi:2009zx}, 
flavor-changing interactions involving the neutral heavy Higgs bosons ($H^0$ and $A^0$) are not suppressed in the alignment limit~\cite{Craig:2013hca,Carena:2013ooa}.
In this limit, characterized by $\cos(\beta - \alpha) \approx 0$, 
the light Higgs boson ($h^0$) mimics the Standard Model Higgs boson, 
$h^0_{\rm SM}$. 
As a result, searches for $t \to c h^0$ and $t \to c H^0$ are complementary, 
given that the relevant flavor-changing couplings are 
$\lambda_{tch} = \rho_{tc} \cos(\beta - \alpha)$, 
$\lambda_{tch} = \rho_{tc} \sin(\beta - \alpha)$, 
and $\lambda_{tcA} = \rho_{tc}$. 
In particular, the decays $t \to c H^0$ and $t \to c A^0$ are directly
sensitive to the FCNH coupling $\rho_{tc}$ in the alignment limit with
$\sin(\beta - \alpha) \approx 1$.

We investigated the discovery potential for such processes with a focus on
$pp\to t \bar{t} \to t c \phi^0 +X,\ \phi^0 = H^0 \ \text{or}\ A^0$
where one top quark decays hadronically and the other through
$t \to c \phi^0$ with subsequent
$\phi^0 \to \tau^+\tau^- \to \tau_{lep} \tau_{had}$ decays at the LHC.
One of the top quarks can become virtual if $m_{H} > m_t$.
The two primary physics backgrounds to this signal are
(a) $t\overline{t} j$ with one top quark decaying hadronically and the
other decaying leptonically, and (b) $t\overline{t} jj$
with both top quarks decaying leptonically.
These can be reduced through the use of the following:
\begin{itemize}
\item[(i)] $\tau$-tagging,
\item[(ii)] collinear approximation to reconstruct $\tau$ decays,
\item[(iii)] invariant mass of tau pairs peaking near the Higgs boson mass
  ($M_{\tau\tau} \approx m_\phi$),
\item[(iv)] for $m_\phi < m_t$,
  the reconstructed top quark mass ($M_{c\tau\tau} \approx m_t$), and
\item[(v)] for $m_\phi < m_t$, the charm quark energy $E_c \approx E_c^*$
  in the top quark rest frame.
\end{itemize}

Based on our event level analysis, we find that the LHC
at $\sqrt{s} = 14$~TeV, with $\mathcal{L} = 3000$~fb$^{-1}$, will be
able to detect this FCNH signal  
for $\tilde{\rho}_{tc} \agt 0.1$ and $m_{\phi} < 170$ GeV.
In the case of degeneracy, $m_{H} \simeq m_{A}$,
the signal cross section and the discovery significance can be doubled.

\section*{Acknowledgments}

This research was supported in part by the U.S. Department of Energy and
the University of Oklahoma.


\end{document}